\documentclass[twocolumn]{aastex63}

\usepackage[caption=false]{subfig}
\usepackage{float}
\usepackage{graphicx}	
\usepackage{amsmath}	
\usepackage{amssymb}	

\received{XXX}
\revised{XXX}
\accepted{XXX}
\submitjournal{ApJ}

\shorttitle{Mini-Neptune Atmospheres}
\shortauthors{Piette \& Madhusudhan}
\graphicspath{{./}{figures/}}

\begin{document}

\title{On the Temperature Profiles and Emission Spectra of Mini-Neptune Atmospheres}

\author[0000-0002-4487-5533]{Anjali A. A. Piette}
\email{ap763@ast.cam.ac.uk}
\affil{Institute of Astronomy, University of Cambridge, Madingley Road, Cambridge CB3 0HA, UK}

\author[0000-0002-4869-000X]{Nikku Madhusudhan}
\email{nmadhu@ast.cam.ac.uk}
\affil{Institute of Astronomy, University of Cambridge, Madingley Road, Cambridge CB3 0HA, UK}



\begin{abstract}
Atmospheric observations of mini-Neptunes orbiting M-dwarfs are beginning to provide constraints on their chemical and thermal properties, while also providing clues about their interiors and potential surfaces. With their relatively large scale heights and large planet-star contrasts, mini-Neptunes are currently ideal targets towards the goal of characterising temperate low-mass exoplanets. Understanding the thermal structures and spectral appearances of mini-Neptunes is important to understand various aspects of their atmospheres, including radiative/convective energy transport, boundary conditions for the interior, and their potential habitability. In the present study, we explore these aspects of mini-Neptunes using self-consistent models of their atmospheres. We begin by exploring the effects of irradiation, internal flux, metallicity, clouds and hazes on the atmospheric temperature profiles and thermal emission spectra of temperate mini-Neptunes. In particular, we investigate the impact of these properties on the radiative-convective boundary and the thermodynamic conditions in the lower atmosphere, which serves as the interface with the interior and/or a potential surface. Building on recent suggestions of habitability of the mini-Neptune K2-18~b, we find a range of physically-motivated atmospheric conditions that allow for liquid water under the H$_2$-rich atmospheres of such planets. We find that observations of thermal emission with JWST/MIRI spectrophotometry can place useful constraints on the habitability of temperate mini-Neptunes such as K2-18~b, and provide more detailed constraints on the chemical and thermal properties of warmer planets such as GJ~3470~b. Our results underpin the potential of temperate mini-Neptunes such as K2-18~b as promising candidates in the search for habitable exoplanets.
\end{abstract}




\section{Introduction}
\label{sec:intro}
Of the thousands of exoplanets known to date, mini-Neptunes are the most common type \citep{Fressin2013,Fulton2017}. Yet, no analogue is known in the solar system. We refer to a mini-Neptune as a planet with a mass and radius smaller than Neptune whose density is too low to be explained by a purely rocky composition \citep[e.g.][]{Rogers2015,Fulton2017}. With properties between those of rocky planets and ice giants, these intermediate objects offer an intriguing window into a new realm of atmospheric, surface and interior processes.

Mini-Neptunes arguably represent optimal targets for atmospheric studies of temperate planets with potentially `habitable' surfaces. By ‘habitable’ we mean thermodynamic conditions at which liquid water is possible on a planetary surface \citep[e.g.,][]{Kasting1993,Meadows2018book} and at which life is known to survive on Earth, including extremophiles, i.e., at temperatures $\lesssim$ 400~K and pressures $\lesssim$~1250 bar \citep[e.g.,][]{Rothschild2001,Merino2019}. On the one hand, temperate rocky planets in the habitable-zone can host surfaces conducive to life \citep[e.g.][]{Yang2013,Koll2016,Wolf2017,Lincowski2018,Meadows2018book} but their small radii and large atmospheric mean molecular weights make observations challenging \citep[][]{Barstow2016,Dewit2018,Lustig-Yaeger2019}. On the other hand, larger Neptune-size ice giants with H$_2$-rich atmospheres are ideal for atmospheric characterisation \citep[e.g.][]{Fraine2014,Wakeford2017} but the high pressures and temperatures in their deep atmospheres preclude habitable conditions. Temperate mini-Neptunes with the right mass and radius offer the potential for habitable conditions below the atmosphere \citep{Madhusudhan2020}, while their hydrogen-dominated atmospheres make them conducive for spectroscopic observations, similarly to ice giants \citep[e.g.,][]{Kreidberg2014a,Chen2017b,Tsiaras2019,Benneke2019a,Benneke2019b}.

Several studies have investigated the effects of various parameters on physical processes in H$_2$-rich atmospheres. For example, varying $T_\mathrm{int}$ changes the location of the radiative-convective boundary and the temperature structure in the lower atmosphere, with implications for chemical mixing and atmospheric circulation \citep{Morley2017a, Carone2019, Thorngren2019}. While atmospheric mixing through eddy diffusion in the radiative zone can homogenise the observable atmosphere (e.g. P$\lesssim$1 bar; \citealt{Moses2011}), convection makes it more efficient to mix material from the deep atmosphere. Similarly, clouds and hazes with strong optical scattering can cool the atmosphere as they reflect incident stellar irradiation \citep[e.g.][]{Morley2015}, which in turn can affect the surface and interior temperatures possible in such planets. The effects of other physical parameters on H$_2$-rich atmospheres of Neptunes/mini-Neptunes have also been explored in previous works, such as metallicity \citep{Spiegel2010} and UV flux \citep{Moses2013b,Miguel2014}. The effects of photospheric clouds and hazes on mini-Neptune atmospheric spectra have also been explored in detail \citep{Howe2012,Morley2013}. Furthermore, \citet{Malik2019} have recently used self-consistent atmospheric models to explore the occurrence of thermal inversions in both hydrogen-rich and higher-metallicity super-Earth atmospheres.

While habitability studies often focus on Earth-like terrestrial planets, this has also been explored for larger planets with H$_2$-rich atmospheres. Since a nominal definition of habitability requires the presence of liquid surface water  \citep[e.g.][]{Meadows2018book}, temperate conditions are required at the planetary surface to allow for habitability. For Earth-like rocky exoplanets, surface temperatures and pressures are regulated by their thin atmospheres dominated by gases such as N$_2$, H$_2$O and CO$_2$, which define the traditional habitable zone \citep[e.g.][]{Kasting1993,Kopparapu2017,Meadows2018a,Meadows2018book}. Recent studies have shown that H$_2$-dominated atmospheres on such planets could also cause a significant greenhouse effect \citep{Stevenson1999,Pierrehumbert2011,Koll2019}, and can extend the traditional habitable-zone out to larger orbital separations. For example, \citet{Pierrehumbert2011} argue for the habitability of such planets out to 1.5 AU for a 3 M$_\oplus$ planet orbiting an early M Dwarf, which could be accessible by microlensing. However, atmospheric characterisation of such planets is challenging due to their larger orbital distances. In this study, we focus on planets with H$_2$-rich atmospheres in close-in orbits which are accessible with transit observations and consider the conditions under which their surfaces could be habitable. 

While past studies of habitability under H$_2$-rich atmospheres have been focused on rocky super-Earths \citep[e.g.][]{Pierrehumbert2011,Seager2013b}, it is becoming evident that certain mini-Neptunes with H$_2$-rich atmospheres may also host habitable surfaces \citep{Madhusudhan2020}. Such mini-Neptunes must have masses and radii, i.e., bulk densities, that can allow a sufficiently low surface pressure for a H$_2$O layer under the H$_2$-rich atmosphere. Extremophiles on Earth are known to survive pressures as high as 1250 bar and temperatures up to 395~K \citep{Merino2019}. Therefore, a mini-Neptune with a surface pressure of $\lesssim$1000~bar could potentially be habitable if its surface temperature is cool enough. \citet{Madhusudhan2020} find that this is the case for the habitable-zone mini-Neptune K2-18~b. They find that the mass, radius and atmospheric properties of K2-18~b are consistent with an interior structure comprising a silicate/Fe core, a H$_2$O layer and a H$_2$-rich atmosphere. The pressure below the atmosphere can be as low as $\sim$1~bar with temperatures $\lesssim$400~K, allowing habitable conditions in the H$_2$O layer underneath. 

K2-18~b may therefore represent a Rosetta Stone for exoplanetary habitability, with the potential for both habitable conditions and detailed atmospheric characterisation with current and future facilities. Of course, unlike K2-18~b, not all mini-Neptunes are expected to host habitable surfaces. Mini-Neptunes that are hotter and with lower bulk densities than K2-18b would lead to significantly higher temperatures and pressures, respectively, below their H$_2$-rich atmospheres that would be too high to be habitable. For example, GJ~1214~b, with similar mass and radius to K2-18~b but with an equilibrium temperature of $\sim$500K, is expected to have super-critical H$_2$O beneath its atmosphere at temperatures unconducive to life \citep{Rogers2010b}. However, for planets such as K2-18b, a wide range of atmospheric parameters may allow for habitable conditions. Indeed, conditions even slightly hotter than K2-18b and a range of surface pressures may allow for habitability, as we explore in this study. K2-18b may, therefore, be the archetype for this class of planets, many more of which may be discovered by current and upcoming surveys. 

In this study, we focus on three primary aspects of mini-Neptunes atmospheres. Firstly, we explore the temperature structures and spectral appearance of mini-Neptunes as a function of key atmospheric parameters, such as incident irradiation, internal flux, metallicity, and cloud/haze properties. In particular, we investigate the interplay between radiative and convective energy transport mechanisms in mini-Neptune atmospheres as a function of these parameters, with important implications for atmospheric composition and dynamics. Secondly, we investigate the implications of atmospheric temperature structures of mini-Neptunes on their surface conditions underneath the atmosphere and assess their potential habitability. Finally, we evaluate the observability of mini-Neptune atmospheres in thermal emission with JWST to characterise both potentially-habitable as well as warmer mini-Neptunes. In particular, we propose a simple metric to identify temperate mini-Neptunes which could potentially host habitable conditions, with a view to guiding follow-up observations. We demonstrate our results on some case studies, including the habitable-zone mini-Neptune K2-18~b and the warmer mini-Neptune GJ~3470~b. 

In what follows, we begin by outlining our atmospheric model in section \ref{sec:model}. In section \ref{sec: results}, we then present a suite of self-consistent $P$-$T$ profiles and emergent spectra exploring the atmospheric parameter space of mini-Neptunes using K2-18~b as a prototype. We further conduct a detailed study of the mini-Neptune K2-18~b in section \ref{sec:case_studies} with the specific goal of assessing the thermodynamic conditions at the base of the atmosphere and the potential for habitable conditions therein. In section \ref{sec:observability} we investigate the observability of mini-Neptunes, both temperate and warmer planets, with JWST in the mid-infrared. We conclude and discuss our findings in section \ref{sec:discussion}.

\section{Atmospheric Model}
\label{sec:model}

We model the atmospheres of mini-Neptunes using an adaptation of the self-consistent atmospheric modeling framework \textsc{GENESIS} (\citealt{Gandhi2017}; also used in \citealt{Piette2020}). \textsc{GENESIS} self-consistently calculates the pressure-temperature ($P$-$T$) profile, the chemical profile and the spectrum of a plane-parallel atmosphere under assumptions of radiative-convective, hydrostatic, local thermodynamic and thermochemical equilibrium. In what follows, we discuss the model considerations and adaptations to \textsc{GENESIS} made in the present study, including radiative transfer, clouds/hazes, and comparisons with similar models in the literature.

\subsection{Energy budget and radiative-convective equilibrium}
\label{sec:radconv}
Applying radiative-convective equilibrium to the model atmosphere involves balancing inward and outward energy transport in each layer of the atmosphere. The total energy budget of the atmosphere can be thought of as coming from two sources: incident irradiation from the host star and an intrinsic flux emanating from within the planet, which represents residual energy from formation/accretion and can also include other heating effects such as tidal heating. The level of incident irradiation can be characterised by the irradiation temperature,
\begin{equation}
\label{eq:teq}
T_\mathrm{irr} = \left(\frac{R_\star}{\sqrt{2}a}\right)^\frac{1}{2}T_\star.
\end{equation}
where $T_\star$ and $R_\star$ are the stellar temperature and radius, respectively, and $a$ is the semi-major axis of the planet. $T_\mathrm{irr}$ is effectively the equilibrium temperature ($T_\mathrm{eq}$) of the dayside atmosphere of the planet assuming zero albedo and that the incident flux is redistributed and reradiated only on the dayside, i.e., no day-night redistribution. 

In our model we assume that the incident irradiation is uniformly redistributed over the day-side plane parallel atmosphere. A significant amount of the incident irradiation is expected to be reflected back in the presence of strong optical scattering (e.g. due to clouds or hazes) as we consider in the present work. A fraction of the remnant irradiation that reaches the deeper atmosphere may be transported to the night-side depending on the location of the atmosphere-interior boundary. For example, in cases where the boundary occurs at low pressures, the day-night redistribution may take place in the interior, e.g., the H$_2$O layer, rather than the atmosphere. We, therefore, do not include any day-night atmospheric redistribution explicitly; we discuss this in section \ref{sec:discussion}.

The internal flux can be characterised by internal temperature, $T_\mathrm{int}$. Values of $T_\mathrm{int}$ can depend on formation mechanisms, mass, composition, internal sources of heating and age, since a planet loses its initial energy from formation over time at a rate determined by its internal and atmospheric properties. Estimates of $T_\mathrm{int}$ can therefore be made using planetary evolution models \citep[e.g.][]{Valencia2013,Lopez2014}. In this work, we explore a wide range of plausible internal temperatures for mini-Neptunes. \citet{Valencia2013} find that for the mini-Neptune GJ~1214~b, $T_\mathrm{int}$ can be as high as 80~K/62~K if it has a water-rich/solar-like composition and a young age of 0.1 Gyr. \citet{Morley2017a} consider higher values of $T_\mathrm{int}$ ($\gtrsim$300~K) for GJ~436b as this planet may be experiencing tidal heating. Conversely, the minimum possible value of $T_\mathrm{int}$ is 0 K. In this work, we explore the range $T_\mathrm{int}=$0-200 K.

Once the energy budget of the atmosphere is set, energy transport can occur in two primary ways: radiative and convective transport. In \textsc{genesis}, convective fluxes are calculated using mixing length theory \citep{Kippenhahn2012} for regions of the atmosphere where the temperature gradient is steeper than the adiabatic gradient. In purely radiative regions of the atmosphere, the requirement of energy balance in a given atmospheric layer can be written as
\begin{displaymath}
\int_{0}^{\infty}\kappa_\nu(J_\nu-B_\nu)d\nu=0
\end{displaymath}
where $\nu$ is frequency, $J_\nu$ is the mean intensity of radiation in that layer, $B_\nu$ is the Planck function corresponding to the temperature in the layer and $\kappa_\nu$ is the absorption coefficient. This form of the equations assumes that the atmosphere is in local thermodynamic equilibrium, and the $\kappa_\nu J_\nu$ and $\kappa_\nu B_\nu$ terms represent the absorption and emission of radiation in the atmospheric layer, respectively. In convective regions of the atmosphere, the convective flux is also added to this equation \citep[see][for more detail]{Gandhi2017}. In order to find a $P$-$T$ profile which satisfies energy balance, \textsc{genesis} then uses Rybicki's method to iteratively solve the equations of radiative-convective equilibrium \citep{Hubeny2014,Gandhi2017}.

\subsection{Radiative transfer}
\label{sec:radtrans}
Given a $P$-$T$ profile and the opacity structure of an atmosphere, the radiation field of the atmosphere can be calculated using the radiative transfer equation. For a given frequency, $\nu$, and atmospheric layer, this equation is
\begin{equation}
    \label{eq:radtrans}
    \mu \frac{\mathrm{d}I_\nu}{\mathrm{d}\tau_{\nu}} = I_\nu - S_\nu,
\end{equation}
where $I_\nu$ is specific intensity at angle $\theta$ to the normal, $\tau_{\nu}$ is the optical depth of the layer, $S_\nu$ is the source function in the layer and $\mu=\cos\theta$. 

This procedure is necessary at two points in the calculation of the atmospheric model: (i) in the iterative solution of radiative-convective equilibrium, and (ii) in order to calculate the emergent spectrum of the planet once a converged, energy-balanced atmospheric model is found. For the solution of radiative-convective equilibrium, a fast radiative transfer solution is ideal as the calculation needs to be performed in each iteration. Furthermore, values of mean intensity are not directly needed from this solution; instead, Eddington factors ($f_\nu=K_\nu/J_\nu$, where $K_\nu=1/2\int_{-1}^{1}\mu^2I_\nu(\mu)\mathrm{d}\mu$) are sufficient for the calculation of radiative equilibrium. On the other hand, calculation of the emergent spectrum requires accurate solutions for the mean intensity and is only calculated once, so need not be as fast. We therefore choose to use the Feautrier method \citep{Feautrier1964} in the iterative solution of radiative-convective equilibrium, and the Discontinuous Finite Element (DFE) method \citep{Castor1992} combined with Accelerated Lambda Iteration (ALI) for the calculation of the spectrum. Both methods are second-order accurate and provide direct solutions to the radiative transfer equation.

The Feautrier method solves radiative transfer under the assumption that the source function is isotropic. This is achieved by recasting equation \ref{eq:radtrans} in terms of the symmetric and anti-symmetric averages of specific intensity, $j_\nu=(I_\nu(\mu)+I_\nu(-\mu))/2$ and $h_\nu=(I_\nu(\mu)-I_\nu(-\mu))/2$, respectively. This results in the Feautrier equation, $\mu^2\frac{\mathrm{d}^2j_\nu}{\mathrm{d}\tau_{\nu}^2}=j_\nu-S_\nu$. This equation is then solved using matrix methods as described in \citet{Hubeny2014}, and yields the desired Eddington factors which are used in the solution of radiative-convective equilibrium.

The DFE solution for radiative transfer divides the atmosphere into plane-parallel layers and solves for the specific intensity at the top and bottom of each layer, $I_d^+$ and $I_d^-$, respectively, for layer $d$. Crucially, a discontinuity between adjacent layers is allowed, i.e. $I_d^+\neq I_{d+1}^-$, and it is this property which allows the method to be second-order accurate \citep{Castor1992}. The specific intensity for layer $d$, $I_d$ is then given by a weighted average of $I_d^+$ and $I_d^-$. This formalism leads to the following recurrence relations for $I_d^+$ and $I_d^-$ (dropping the $\nu$ subscript for clarity):
\begin{align*}
a_d I_{d+1}^- &= 2I_{d}^- + \Delta\tau_{d+1/2}S_{d} +b_dS_{d+1} \\
a_d I_{d}^+ &= 2(\Delta \tau_{d+1/2}+1)I_{d}^- +b_d S_{d} - \Delta\tau_{d+1/2}S_{d+1/2},   
\end{align*}
where $a_d=\Delta\tau_{d+1/2}^2+2\Delta\tau_{d+1/2}+2$, $b_d=\Delta\tau_{d+1/2}(\Delta\tau_{d+1/2}+1)$ and $\Delta\tau_{d+1/2}=(\tau_{d+1}-\tau_d)/|\mu|$. When scattering is present, this method must be used iteratively, which can be done using ALI. The general principle of ALI methods is to write a matrix equation such as $I=\Lambda [S]$ as an iterative process, i.e. $I^{\mathrm{new}} = \Lambda^\ast[S^{\mathrm{new}}]+(\Lambda-\Lambda^\ast)[S^\mathrm{old}]$ where $\Lambda^\ast$ is an approximate operator chosen to minimise computation time and maximise convergence rate. Details of the implementation of ALI for the DFE method can be found in \citet{Hubeny2017}.

\begin{figure*}
    \centering
    \includegraphics[width=0.8\textwidth]{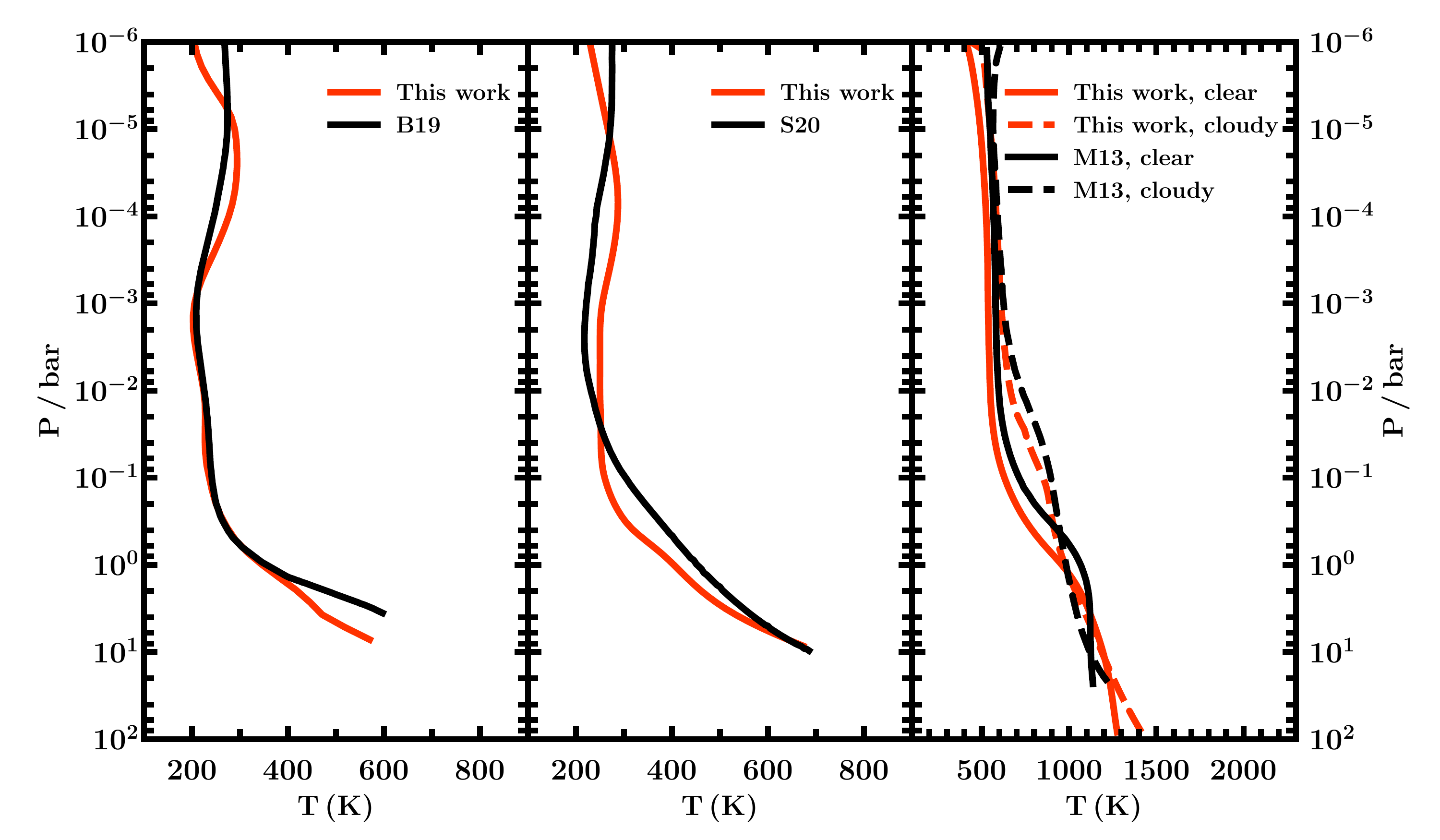}
    \caption{Comparisons with model pressure-temperature profiles of mini-Neptunes in the literature. Left panel: comparison to the model for K2-18~b from figure 5 of \citet{Benneke2019b} (B19). Center panel: comparison to the K2-18~b solar-metallicity model from figure 11 of \citet{Scheucher2020} (S20, case 6 in their table 6). Right panel: comparison to clear (solid lines) and cloudy (dashed lines) models for GJ~1214~b from \citet{Morley2013} (M13). See section~\ref{sec:modelcomp} for details.} 
    \label{fig:modelcomp}
\end{figure*}

\subsection{Chemistry, opacities and clouds/hazes}
\label{sec:opacity}
In this work, we consider opacity due to species in gas phase, hazes and clouds. The gas-phase species we include here are the main volatiles expected in hydrogen-rich atmospheres: H$_2$O, CH$_4$, NH$_3$, CO, CO$_2$, HCN and C$_2$H$_2$ \citep{Madhusudhan2012, Moses2013a}. We calculate their equilibrium abundances self-consistently with the $P$-$T$ profile using the analytic method of \citet{Heng2016} in each iteration of the radiative-convective equilibrium solver. We also include opacity due to H$_2$-H$_2$ and H$_2$-He collision-induced absorption (CIA). The cross sections we use for these species are calculated from the HITEMP, HITRAN and ExoMol line list databases using the methods of \citet{Gandhi2017} (H$_2$O, CO and CO$_2$: \citealt{Rothman2010}, CH$_4$: \citealt{Yurchenko2013,Yurchenko2014a}, C$_2$H$_2$: \citealt{Rothman2013,Gordon2017}, NH$_3$: \citealt{Yurchenko2011}, HCN: \citealt{Harris2006,Barber2014}, CIA: \citealt{Richard2012}). 

In our models we also consider the phase transition of gaseous H$_2$O in the atmosphere into both the liquid and ice phases. When H$_2$O condenses to the liquid phase, we remove it from the atmosphere as precipitation. This `rainout' happens if the H$_2$O mixing ratio is greater than the local saturation vapour pressure. In this case, we assume that any H$_2$O in excess of the saturation level rains out, and the remaining atmospheric H$_2$O vapour pressure is equal to the saturation vapour pressure in these regions. On the other hand, in regions of the atmosphere where the temperature is below the freezing point, the H$_2$O is removed entirely from the gas phase and instead included as an ice cloud as described below.

We also include a simple prescription for hazes in our models. We assume that the haze is homogeneously distributed in the atmosphere and its opacity is of the form of an enhanced H$_2$ Rayleigh scattering; i.e. H$_2$ Rayleigh scattering boosted by a multiplicative factor. This allows us to explore the general effects of haze on model $P$-$T$ profiles with a simple parameterisation. In order to compare the strength of our model hazes to more detailed studies, we look to the models of \citet{Howe2012} for GJ~1214~b. One of the best fit models presented by \citet{Howe2012} includes a tholin haze of density 100 cm$^{-3}$ and particle size 0.1 $\mu$m, whose extinction has a $\lambda^{-4}$ dependence at wavelengths $<$1 $\mu$m. The scattering opacity of this haze is equivalent to $\sim$1000$\times$ the scattering opacity of H$_2$ Rayleigh scattering at 1~mbar and 300~K. In section \ref{sec: results}, we therefore explore haze opacities in the range 0-10,000$\times$ H$_2$ Rayleigh scattering.

Our prescription for clouds involves cloud decks whose particle abundances decay exponentially with altitude above the cloud base given a cloud scale height, analogous to the condensate profiles of \citet{Ackerman2001}. Rather than parameterising the cloud scale height according to parameters such as $f_\mathrm{rain}$ as in \citet{Ackerman2001}, we choose for simplicity a nominal cloud scale height of 1/3 of an atmospheric scale height, similar to Jovian ammonia clouds (e.g., \citealt{Carlson1994}; \citealt{Brooke1998}, see also \citealt{Ackerman2001}). The base of the deck is fixed at a given pressure, which in sections \ref{sec:case_studies} and \ref{sec:observability} we choose such that the temperature at the base of the cloud approximately coincides with the condensation temperature of the cloud species \citep[e.g. see condensation curves of][]{Morley2012}. In section \ref{sec: results}, however, we independently explore the effect of cloud location on the $P$-$T$ profile and therefore do not consider condensation temperature. We nominally assume a KCl composition to represent the effects of salt clouds, and also explore the effects of water ice clouds. In section \ref{sec:case_studies}, we include KCl, ZnS and ice clouds wherever they are thermodynamically expected to occur. 

We vary cloud opacity in our models by varying the abundance of the condensate species (i.e. varying metallicity). At the base of each cloud, we assume that all of the cloud species is in the condensed phase. For the salt clouds, the abundance of cloud particles is therefore determined by the abundance of the least-abundant element in the condensate species (e.g. Cl for KCl) and the particle size. Cloud particle sizes are known to vary depending on a range of physical factors which have been explored in detail in several works \citep[e.g.][]{Ackerman2001,Morley2014}. For simplicity, throughout this work, we assume a nominal modal particle size of 0.33~$\mu$m  for the salt clouds and use the cloud extinction cross sections given by \citet{Pinhas2017}. For ice clouds, the particle abundance at the base of the cloud depends on the H$_2$O abundance and the particle size (which can vary between models, see section \ref{sec:case_studies}). We use the extinction cross sections for water ice from \citet{Budaj2015}. 

Above an ice cloud deck, the abundance of gaseous H$_2$O can depend on the presence of vertical mixing. For our model parameter exploration in section \ref{sec: results}, we freeze out all H$_2$O above the base of the water ice cloud deck (i.e. assuming a cold trap and no vertical mixing) to provide a uniform comparison between models with different ice cloud abundance, which is a free parameter. In reality, however, if a thermal inversion occurs it is possible that the temperature profile above the cloud deck is hotter than the freezing point. In such cases, it is conceivable that water vapour may be present above the cloud deck due to vertical mixing in the atmosphere; e.g., analogous to condensable species being lofted to upper regions in irradiated atmospheres \citep[e.g.,][]{Parmentier2013}. For example, if the temperature at 0.1~bar is 250~K and the temperature at 10$^{-3}$~bar is 300~K, H$_2$O will be frozen out at 0.1~bar but H$_2$O vapour could still be present at 10$^{-3}$~bar given sufficient vertical mixing. Therefore, we allow for this possibility when modeling the specific case studies in sections \ref{sec:case_studies} and \ref{sec:observability}, i.e., H$_2$O is only frozen out when it is expected to be in the ice phase, allowing it to come back to vapor phase higher up in the atmosphere if the temperature there is above the freezing point.  

\subsection{System parameters}
\label{sec:params}

In section \ref{sec: results}, we apply our model to a generic mini-Neptune to qualitatively asses the impacts of various atmospheric parameters on its $P$-$T$ profile and spectrum. The planetary and host star parameters we choose to use for this test case are based on those of K2-18~b \citep{Foreman-Mackey2015, montet2015}, and we use the values given by \citet{Cloutier2019} and \citet{Benneke2019b}. For the planetary parameters, we assume a radius of 2.61$R_\oplus$ and log gravity of 3.094 (in cgs). For the stellar parameters, we assume a radius of 0.4445$R_\odot$, log gravity of 4.836 (in cgs), temperature of 3457~K and an [Fe/H] metallicity of 0.12. We use a Kurucz stellar model for the stellar spectrum in our models \citep{Kurucz1979,Castelli2003}. 

In section \ref{sec:case_studies}, we calculate models specifically for K2-18~b using the same parameters as described above, and using its orbital separation to determine the irradiation temperature based on parameters from \citet{Benneke2019b}. For K2-18~b, $T_\mathrm{irr}$=332~K. We further estimate the internal temperature of K2-18~b based on existing estimates for GJ~1214~b \citep{Charbonneau2009}, which has a similar mass and radius to K2-18~b \citep{Cloutier2019,Benneke2019b}. \citet{Valencia2013} find that for ages between $\sim$1-10 Gyr, considering both solar and water-rich atmospheric compositions, GJ~1214~b has an internal temperature in the range $\sim$25-50 K. We therefore calculate models for these two end-member values of $T_\mathrm{int}$.

We also calculate models for GJ~3470~b \citep{Bonfils2012} in section \ref{sec:case_studies}, a planet with radius comparable to Neptune but of lower mass. For this, we use planetary and stellar properties from \citet{Awiphan2016}. The planetary radius and log gravity are 4.57$R_\oplus$ and 2.81 (in cgs), respectively, with a mass of 13.9 $M_\oplus$, and semi-major axis of 0.0355~au. For the stellar parameters we use a radius of 0.547$R_\odot$, log gravity of 4.695 (in cgs), effective temperature of 3600~K and an [Fe/H] metallicity of 0.2. The irradiation temperature of GJ~3470~b given these system parameters is 812~K. For the internal temperature, we use an intermediate value of 30~K \citep{Valencia2013}.

In section \ref{sec:modelcomp}, we also model GJ~1214~b for comparison with previous models in the literature. For these models, we use the stellar and planetary parameters of GJ~1214/GJ~1214~b from \citet{Harpsoe2013}. The planetary radius and log gravity are 2.85$R_\oplus$ and 2.88 (in cgs), respectively, and the semi-major axis is 0.0141~au. For the star we use a radius of 0.216$R_\odot$, log gravity of 4.944 (in cgs), effective temperature of 3026~K and an [Fe/H] metallicity of 0.39.

The \textsc{genesis} models described above can be used to generate atmospheric models with arbitrary pressure and wavelength resolution. We choose to use 10000 wavelength points uniformly distributed between 0.4 and 50 $\mu$m and 100 depth layers log-uniformly distributed between pressures of $10^3$ and $10^{-6}$ bar.

\subsection{Model comparison}
\label{sec:modelcomp}

\begin{figure*}
    \centering
    \includegraphics[width=\textwidth]{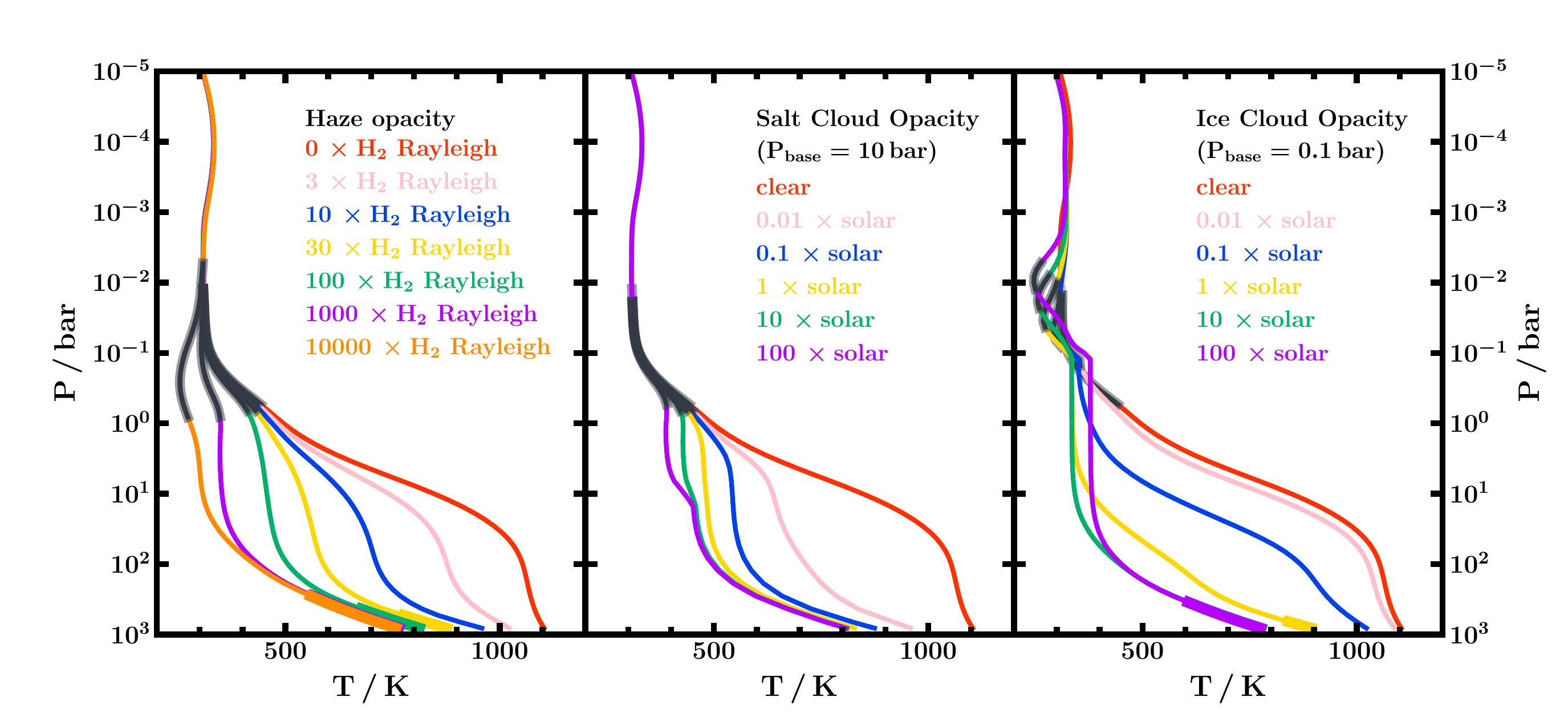}
    \caption{$P$-$T$ profiles for model atmospheres with varying haze and cloud opacities. Left panel shows effects of varying haze, which is included homogeneously throughout the atmosphere
    and implemented using an enhanced Rayleigh scattering prescription (see section~\ref{sec:opacity}). Center and right panels show the effects of a KCl/water ice cloud deck, with the abundance of the cloud condensate parametrised by the metallicity, as described in section~\ref{sec:opacity}. For the salt/water ice clouds, the base of the cloud deck is at a pressure of 10 bar/0.1 bar and the modal particle size is 0.33~$\mu$m/4~$\mu$m, respectively. We use a nominal cloud scale height of 1/3 of an atmospheric scale height. For each model, $T_\mathrm{int}$=40~K, $T_\mathrm{irr}$=350~K and the metallicity of the gaseous species in the atmosphere is 1$\times$solar. In each panel, the red profile denotes a clear atmosphere. Convective regions are shown by bold lines, and photospheres are shown by bold dark grey lines (see section \ref{sec:conv})}
    \label{fig:PT_clouds}
\end{figure*}

\begin{figure*}
    \centering
    \includegraphics[width=\textwidth]{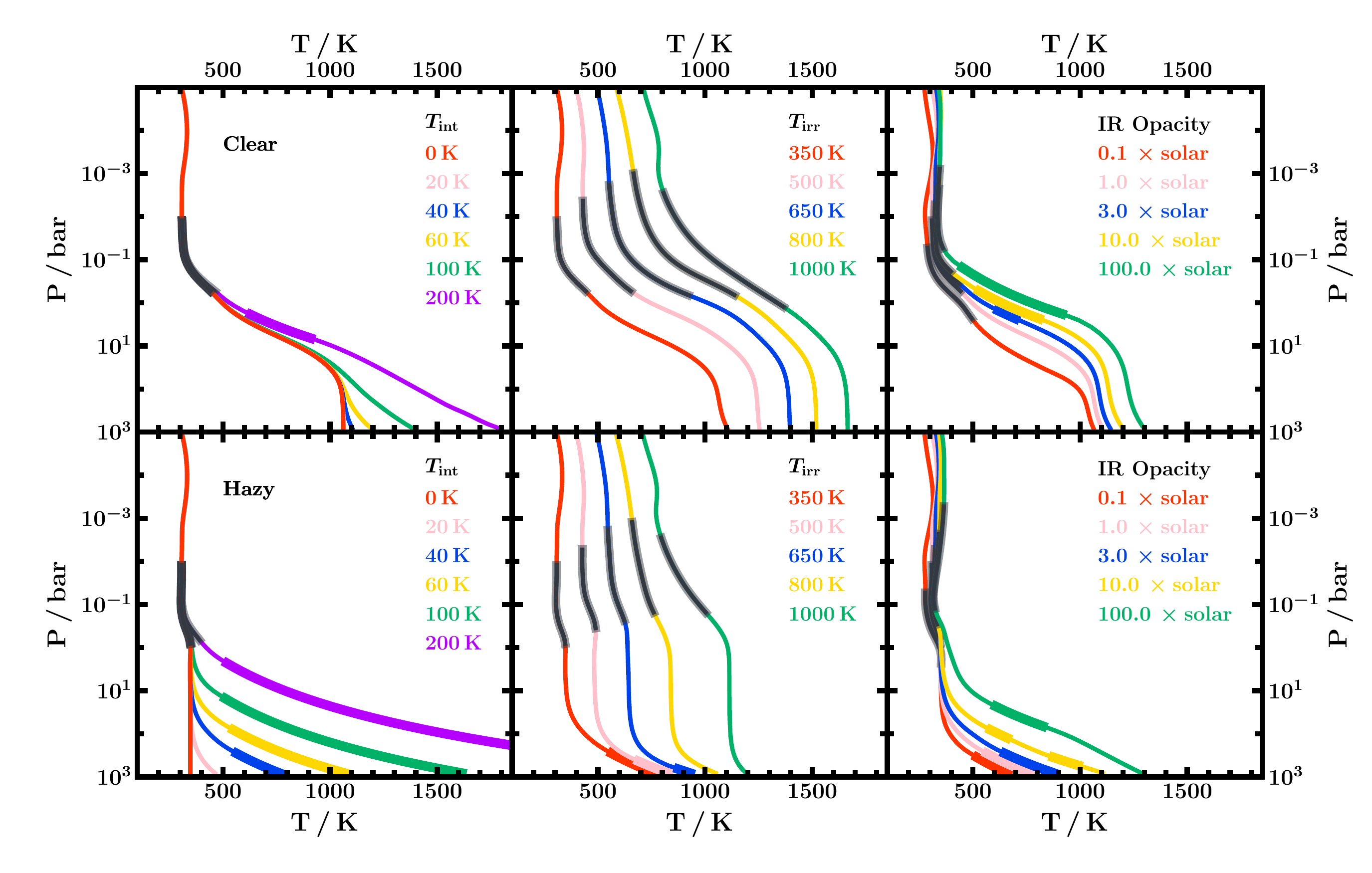}
    \caption{Self-consistent atmospheric $P$-$T$ profiles across a range of internal temperatures, irradiation temperatures and infrared opacities (parameterised by metallicity relative to solar abundances). Top three panels show $P$-$T$ profiles for a clear atmosphere while lower panels include haze equivalent to 1000$\times$ H$_2$ Rayleigh scattering. The fiducial values used for $T_\mathrm{int}$, $T_\mathrm{irr}$ and infrared opacity are 40~K, 350~K and 1$\times$~solar metallicity, respectively. Bold line segments indicate convective regions and photospheres are shown by bold dark grey lines (see section \ref{sec:conv}).}
    \label{fig:PT_main}
\end{figure*}

We compare our self-consistent forward model to three other examples of mini-Neptune models in the literature: \citet{Benneke2019b} and \citet{Scheucher2020} for K2-18b, and \citet{Morley2013} for GJ~1214~b. We use the same input parameters and model set-up as described in these works. The internal temperature is not specified for these models, so we choose to use $T_\mathrm{int}$=100~K as this results in good agreement.

We first reproduce the model $P$-$T$ profile of K2-18~b from figure 5 of \citet{Benneke2019b}. We include opacity due to H$_2$O vapour, H$_2$-H$_2$ and H$_2$-He CIA and Rayleigh scattering due to H$_2$. Following \citet{Benneke2019b}, we assume an albedo of 0.3. We further assume uniform day-night redistribution of the incident flux, i.e., 50\% of the incident irradiation, minus the reflected component, remains on the dayside. We use a H$_2$O abundance of 40$\times$ the expected abundance for a solar composition in thermochemical equilibrium, i.e. approximately the best fitting value found by \citet{Benneke2019b} (their Fig. 4), which they use to calculate their model $P$-$T$ profile. We also use the planetary and stellar parameters for K2-18~b/K2-18 given by \citet{Benneke2019b}, which are also listed in section \ref{sec:params}. The left panel of figure \ref{fig:modelcomp} shows the model from \citet{Benneke2019b} as well as our reproduction, which agrees very closely. 

Note that the $P$-$T$ profile of \citet{Benneke2019b} does not appear to explicitly include the effects of water ice clouds, despite having temperatures below the freezing point of H$_2$O. Instead, an albedo of 0.3 was assumed to remove the corresponding amount of incident flux at the top of the atmosphere. Our reproduction is, therefore, also cloud-free and assumes the same albedo treatment. We explore cloudy and hazy models of K2-18~b in section \ref{sec:case_studies}.

We further reproduce the solar-metallicity $P$-$T$ profile for K2-18~b from figure~11 of \citet{Scheucher2020} (case 6 in their table 6). Their model assumes equilibrium chemical abundances for an isotherm at 320~K, and this chemistry is kept fixed for the calculation of the $P$-$T$ profile. We therefore include opacity due to H$_2$O, CH$_4$ and NH$_3$ in our reproduction (the dominant carriers of O, C and N, respectively, at this temperature) assuming fixed, constant-with-depth abundances corresponding to the equilibrium abundances expected at 320~K (i.e. mixing ratios of $10^{-3}$, $10^{-3.3}$ and $10^{-3.9}$ for H$_2$O, CH$_4$ and NH$_3$, respectively; see e.g., \citealt{Woitke2018}). We also include opacity due to H$_2$-H$_2$ and H$_2$-He CIA as well as H$_2$ Rayleigh scattering. Following \citet{Scheucher2020}, we do not include any clouds or hazes in this model. We also assume uniform day-night energy redistribution. We also use the planetary and stellar parameters for K2-18~b/K2-18 given in section \ref{sec:params}. Both the model from \citet{Scheucher2020} and our reproduction of this are shown in the center panel of figure \ref{fig:modelcomp}.

Our model and that of \citet{Scheucher2020} agree closely. Any differences between them may be due to differences in the treatment of radiative transfer and radiative-convective equilibrium, e.g., \citet{Scheucher2020} do not self-consistently consider convective flux. We further note that this $P$-$T$ profile enters the ice phase of H$_2$O and should therefore include the presence of water ice clouds. We find in section \ref{sec:case_studies} that assessing the habitability of K2-18~b requires consideration of optical opacity in the atmosphere, e.g., clouds and/or hazes. As such, the cloud- and haze-free models of \citet{Scheucher2020} are limited in assessing the habitability of this planet. 

We also use our model to reproduce a clear and a cloudy $P$-$T$ profile for GJ~1214~b from figure 1 of \citet{Morley2013}. We model a 1$\times$solar clear atmosphere and a 50$\times$ solar cloudy atmosphere, both assuming that incident irradiation is redistributed on the dayside only (i.e. corresponding to the hotter $P$-$T$ profiles in figure 1 of \citealt{Morley2013}). In both cases, we include opacity due to all of the volatile species discussed in section \ref{sec:opacity}. In the cloudy model, we also include KCl and ZnS clouds with base pressures of 0.025 and 0.158 bar, respectively. This is where the \citet{Morley2013} model crosses the condensation curves for each of these species and where the cloud bases are positioned in their model. We use a modal particle size of 38.6 $\mu$m for these clouds. Since the effective temperature of GJ~1214 is cooler than 3500~K, we are not able to use a spectral model from the Kurucz library as described above (which includes models with $T_\mathrm{eff}\geq$3500~K) and instead use a model from the PHOENIX library \citep{Phoenix2013} with the closest stellar parameters to GJ~1214, i.e. $T_\mathrm{eff}$=3000~K, log($g$/cgs)=5.0 and [Fe/H]=0.5. Our models and those of \citet{Morley2013} are shown in the right panel of figure \ref{fig:modelcomp}. The models show good agreement and differ by $\lesssim$100~K, which may be due to differences in the stellar spectrum used.

\section{Effects of Atmospheric Parameters}
\label{sec: results}

\begin{figure*}
    \centering
    \includegraphics[width=\textwidth]{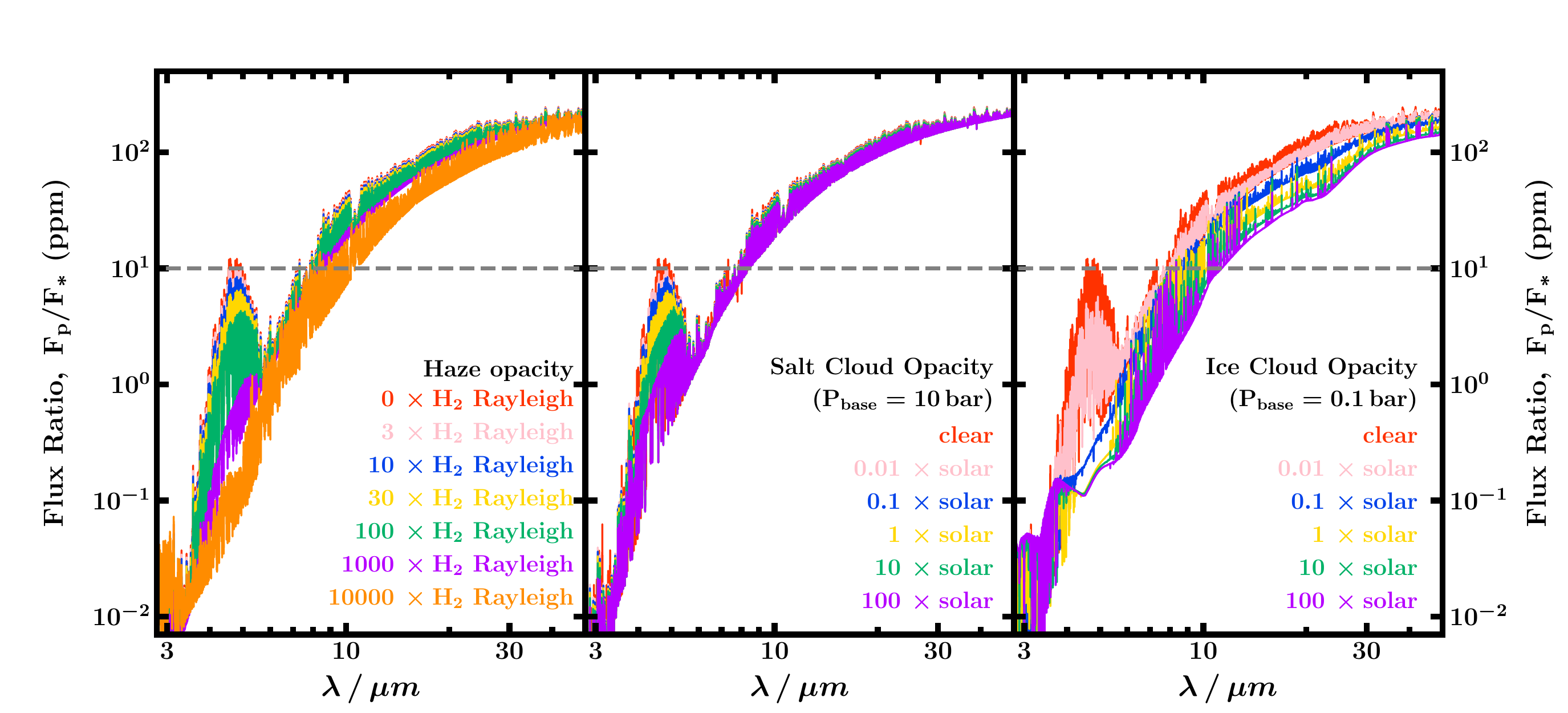}
    \caption{Model thermal emission spectra corresponding to the $P$-$T$ profiles in figure \ref{fig:PT_clouds}. For each model, $T_\mathrm{int}$=40~K, $T_\mathrm{irr}$=350~K and the metallicity of the gaseous species in the atmosphere is 1$\times$solar. We vary haze, salt cloud and water ice cloud opacities in the left, centre and right panels, respectively. The horizontal dashed grey lines show the 10~ppm level, which may be considered an optimistic precision achievable with JWST.}
    \label{fig:spec_clouds}
\end{figure*}

\begin{figure*}
    \centering
    \includegraphics[width=\textwidth]{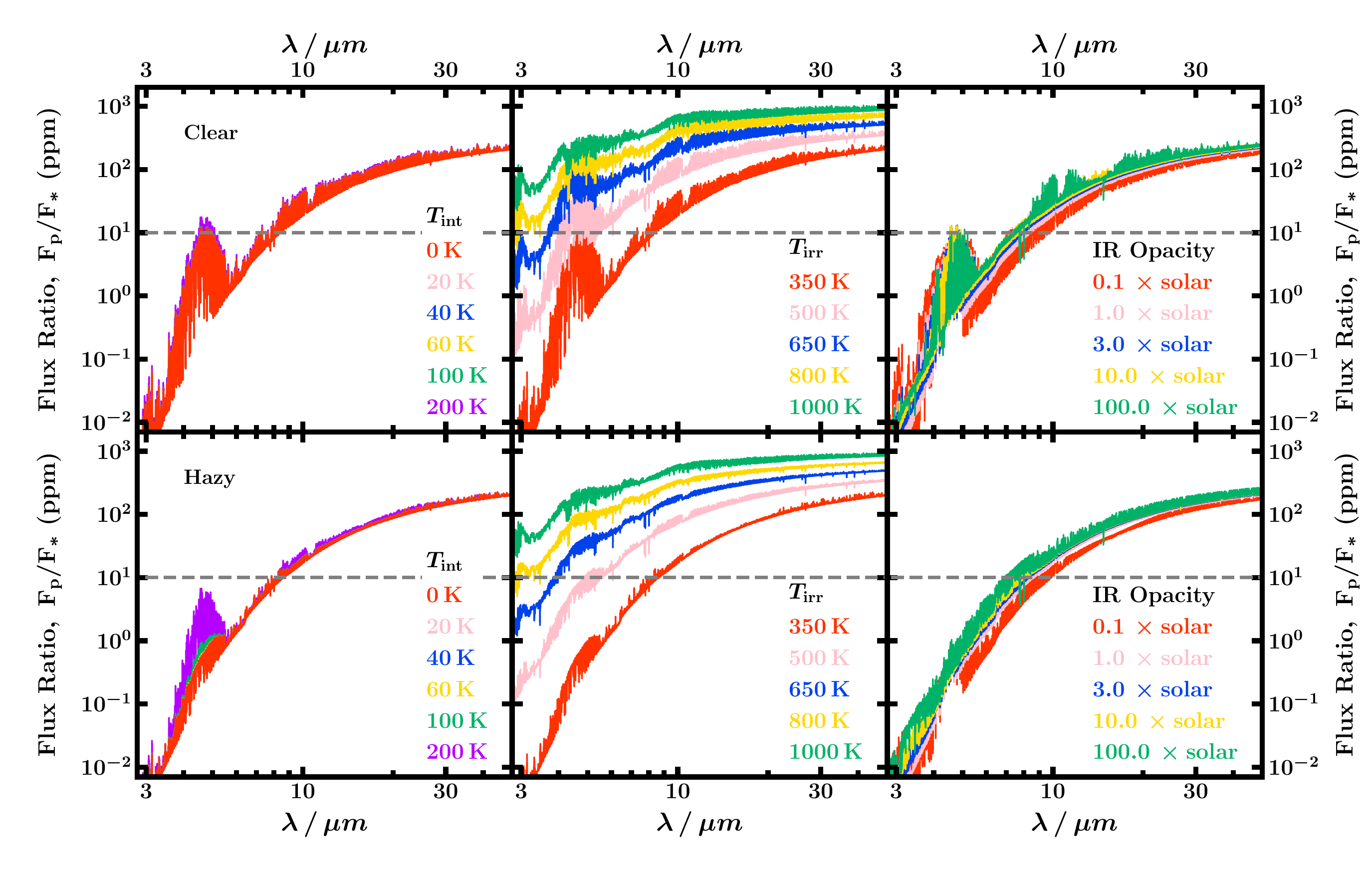}
    \caption{Model thermal emission spectra corresponding to the $P$-$T$ profiles in figure \ref{fig:PT_main}, for varying $T_\mathrm{int}$, $T_\mathrm{irr}$ and infrared opacity (left, centre and right columns, respectively). The fiducial values used for $T_\mathrm{int}$, $T_\mathrm{irr}$ and infrared opacity are 40~K, 350~K and 1$\times$~solar metallicity, respectively. Top three panels show spectra for a clear atmosphere while lower panels include haze equivalent to 1000$\times$ H$_2$ Rayleigh scattering. The horizontal dashed grey lines show the 10~ppm level, which may be considered an optimistic precision achievable with JWST.}
    \label{fig:spec_main}
\end{figure*}

The temperature profiles of mini-Neptune atmospheres are determined by a range of atmospheric properties, including the internal flux, irradiation, and the opacity structure of the atmosphere, which are, therefore, key to understanding the various processes discussed above. In this section, we investigate the effects of these properties on the temperature structures and thermal emission spectra of mini-Neptune atmospheres. In particular, we explore the effects of internal temperature, irradiation, infrared opacity and cloud/haze properties, discussing how they impact the observability and physical processes of these atmospheres as well as consequences for modelling the interiors of mini-Neptunes. 

We consider a fiducial model with $T_\mathrm{irr}$=350~K, $T_\mathrm{int}$=40~K and 1$\times$solar metallicity, and explore models in the range $T_\mathrm{irr}$=350-1000~K, $T_\mathrm{int}$=0-200~K and metallicities of 0.1-100$\times$solar. We present both clear and hazy/cloudy models, exploring haze opacities up to 1000$\times$ Rayleigh scattering (according to the parameterisation discussed in section \ref{sec:opacity}) and salt/water ice clouds with up to 100$\times$solar metallicity. We discuss the $P$-$T$ profiles and thermal emission spectra of these models in sections \ref{sec:conv} and \ref{sec:spectra}, respectively.

\subsection{Temperature profiles and energy transport}
\label{sec:conv}
We begin by exploring how $T_\mathrm{int}$, $T_\mathrm{irr}$, metallicity, cloud/haze opacity and cloud type affect the thermal profiles of mini-Neptunes. Energy transport in a planetary atmosphere is governed by the boundary conditions at the top and bottom of the atmosphere, characterised by $T_\mathrm{irr}$ and $T_\mathrm{int}$, respectively, and by the opacity profile which lies between them. In the upper, low-opacity, regions of the atmosphere, energy transport is primarily radiative, an effect which is enhanced by strong incident irradiation. In the deeper, high-opacity layers, convective transport begins to dominate. Where the transition between the two regimes - the radiative-convective boundary - occurs depends on several factors including incident and intrinsic flux as well as optical and infrared opacity. Depending on the location of the radiative-convective boundary, the presence of convection can in turn impact atmospheric mixing and, therefore, the chemical homogeneity of the atmosphere.

Here, we investigate how the parameters listed above impact the thermal profile and energy transport mechanisms of an atmosphere by generating self-consistent $P$-$T$ profiles, as described in section \ref{sec:model}, and independently varying each parameter in turn. Figure \ref{fig:PT_clouds} shows model $P$-$T$ profiles for which we vary the haze opacity (left panel) or cloud opacity (centre and right panels). In the centre and right panels, we place a KCl cloud at 10~bar and a H$_2$O ice cloud at 0.1~bar, respectively, to test the effects of different types of clouds. We test the effects of $T_\mathrm{int}$, $T_\mathrm{irr}$ and metallicity in figure \ref{fig:PT_main}, for both a clear atmosphere (upper panels) and a hazy one (lower panels). For each $P$-$T$ profile, we also show the 1-30~$\mu$m photosphere, smoothed by a Gaussian of width 0.1~$\mu$m, to represent the pressures and temperatures probed by low-resolution infrared thermal emission observations. In what follows, we describe the effects of each of these parameters in turn on the atmospheric temperature profile.

\subsubsection{Effect of clouds/hazes}
\label{sec:hazecloud}

Figure \ref{fig:PT_clouds} shows the effects of hazes, high-altitude ice clouds and deeper salt clouds on the thermal profile of a mini-Neptune. Here the cloud/haze properties are varied according to the prescriptions described in section~\ref{sec:model}. For example, in the center and right panels the abundance of cloud species is varied according to the metallicity specified, keeping the cloud scale heights and locations fixed. While in reality the cloud base location is driven by its condensation temperature, in this section we choose to only vary cloud opacity in order to independently demonstrate the structural effects that this has on the $P$-$T$ profile. Note that, for the purpose of demonstration here, the abundance of water ice particles is varied independently of the gaseous H$_2$O abundance. The models in figure \ref{fig:PT_clouds} all assume $T_\mathrm{int}$=40~K, $T_\mathrm{irr}$=350~K and solar abundances of the gaseous species listed in section \ref{sec:model}. 

Clouds and hazes both provide optical opacity which scatters incident irradiation and can therefore cool the atmosphere. This is shown clearly in the left and centre columns of figure \ref{fig:PT_clouds}, as increasing the haze/KCl cloud abundance results in a cooler temperature profile. Ice clouds can also cool the atmosphere (right column of figure \ref{fig:PT_clouds}), though very high opacity clouds can also warm the atmosphere by intercepting outgoing flux \citep{Morley2014}. Both the photospheric temperature and the temperature at deeper pressures are affected by these effects, meaning that clouds and hazes are important components in understanding both the spectra and interiors of mini-Neptunes. Figure \ref{fig:PT_clouds} also shows that stronger cloud/haze opacity typically results in a more isothermal temperature profile below the photosphere. This can impact the habitability of the planet as cooler, more habitable temperatures are maintained to higher pressures where a surface may occur. We discuss this further in section \ref{sec:case_studies}. 

The location of the radiative-convective boundary is also affected by the presence of clouds/hazes. As we discuss in section \ref{sec:TeqTint}, the boundary between the radiative and convective regions of the atmosphere is dependent on the incident irradiation, with weaker irradiation resulting in a shallower radiative-convective boundary. By scattering incident irradiation, clouds/hazes also raise this boundary to lower pressures. For example, in figure \ref{fig:PT_clouds}, only the models with the strongest haze/cloud opacities have a radiative-convective boundary shallower than 1000 bar (i.e. the edge of the computational domain), and the boundary occurs at lower pressures for models with higher haze/cloud opacity.

\subsubsection{Irradiation vs internal flux}
\label{sec:TeqTint}
As the boundary conditions at the top and bottom of the atmosphere, irradiation and internal flux compete in determining the location of the radiative-convective boundary. A high $T_\mathrm{irr}$ results in a larger region of the atmosphere being dominated by radiative transport, pushing the radiative-convective boundary deeper. Conversely, a hotter $T_\mathrm{int}$ pushes this boundary higher up. These effects can be seen in figure \ref{fig:PT_main}, and are strongest for the hazy models in the lower left and lower centre panels. As $T_\mathrm{irr}$ decreases the temperature gradient at 1000~bar gradually becomes steeper, transitioning from almost isothermal at $T_\mathrm{irr}=$1000~K to an adiabatic gradient at $T_\mathrm{irr}\leq$650 K, when the radiative-convective boundary occurs within the computational domain. 

As expected, $T_\mathrm{int}$ has the opposite effect. Models with higher $T_\mathrm{int}$ show significantly shallower radiative-convective boundaries, with the shallowest within this parameter space at $\sim$ 1 bar for the model with $T_\mathrm{int}=200$~K. However, such a high value of $T_\mathrm{int}$ may be unlikely for a mini-Neptune \citep[e.g.][]{Valencia2013}, though high $T_\mathrm{int}$ values have been considered for planets which may be affected by tidal heating \citep[e.g. GJ~436b,][]{Morley2017a}. For models with $T_\mathrm{int}\leq$100 K the radiative-convective boundary is deeper than 10 bar. This suggests that intrinsic heat in mini-Neptunes may be insufficient to mix their atmospheres up to shallow pressures through convection unless they have significant haze/cloud opacities. Instead, eddy diffusion in the radiative regime may be a more likely mechanism for vertical mixing.

While $T_\mathrm{irr}$ and $T_\mathrm{int}$ have strong effects on the radiative-convective boundary when haze is present, in the haze-free models their influence is reduced at the pressures investigated here. The upper left and centre panels in figure \ref{fig:PT_main} show that none of the $T_\mathrm{irr}$ or $T_\mathrm{int}$ values explored result in a radiative-convective boundary shallower than 1000 bar for a clear model. Though the larger values of $T_\mathrm{int}$ explored do result in a higher temperature gradient at high pressures, this suggests that for the clear models in the parameter space explored here (i.e. $T_\mathrm{irr} \gtrsim$350~K), incident irradiation and radiative transport dominate the atmosphere up to pressures of at least 1000 bar.

Aside from the radiative-convective boundary, irradiation and internal flux also have significant effects on the temperature profile in general. As expected, a hotter $T_\mathrm{irr}$ translates the temperature profile to higher temperatures, and consequently the photospheric temperature is extremely sensitive to this parameter. In contrast, $T_\mathrm{int}$ only affects the deepest regions of the atmosphere and does not affect the photosphere unless it is extremely high \citep[e.g.][]{Morley2017a}. The left and centre columns of figure \ref{fig:PT_main} also show that the temperatures at high pressures are strongly dependent on both $T_\mathrm{irr}$ and $T_\mathrm{int}$, meaning that these properties are important to consider when using such models as boundary conditions for internal structure models.

\subsubsection{Effect of infrared opacity}
\label{sec:metallicity}
Infrared opacity is an important factor in determining energy transport, as it intercepts outgoing planetary flux. For planets with a cool stellar host whose spectrum peaks in the near-infrared, such as K2-18, the infrared opacity can also absorb incident irradiation. As a result, higher abundances of these infrared absorbers result in a hotter temperature profile. This can be seen in the right column of figure \ref{fig:PT_main}, which shows $P$-$T$ profiles for different infrared opacities, characterised by metallicity relative to solar elemental abundances. These $P$-$T$ profiles also show that metallicity predominantly affects deeper regions of the atmosphere; the profiles are similar at low pressures and begin to diverge at $P\gtrsim$0.1 bar. At these higher pressures, models with higher metallicity also exhibit convective zones as the increase in opacity reduces the efficiency of radiative transport. We also note that for models with very high IR opacity (e.g., $\gtrsim$ 100$\times$ solar), the H$_2$O mixing ratio in the lower atmosphere can be close to saturated. We consider the rainout which can result from super-saturation in sections~\ref{sec:case_studies} and \ref{sec:observability}. 

Since high infrared opacities result in hotter temperatures deep in the atmosphere, this property is an important factor both for boundary conditions in internal structure models and for considerations of habitability. However, the photosphere is not strongly affected by changes in infrared opacity. This is because an increase in infrared opacity effectively translates the $P$-$T$ profile to lower pressures, while the photosphere of a higher-opacity model also occurs at lower pressures \citep{Spiegel2010}. As a result, the photospheres of the models in the right column of figure \ref{fig:PT_main} probe similar temperatures despite the differences in their infrared opacities.

\begin{figure*}
    \centering
        \subfloat{\includegraphics[width=0.48\textwidth]{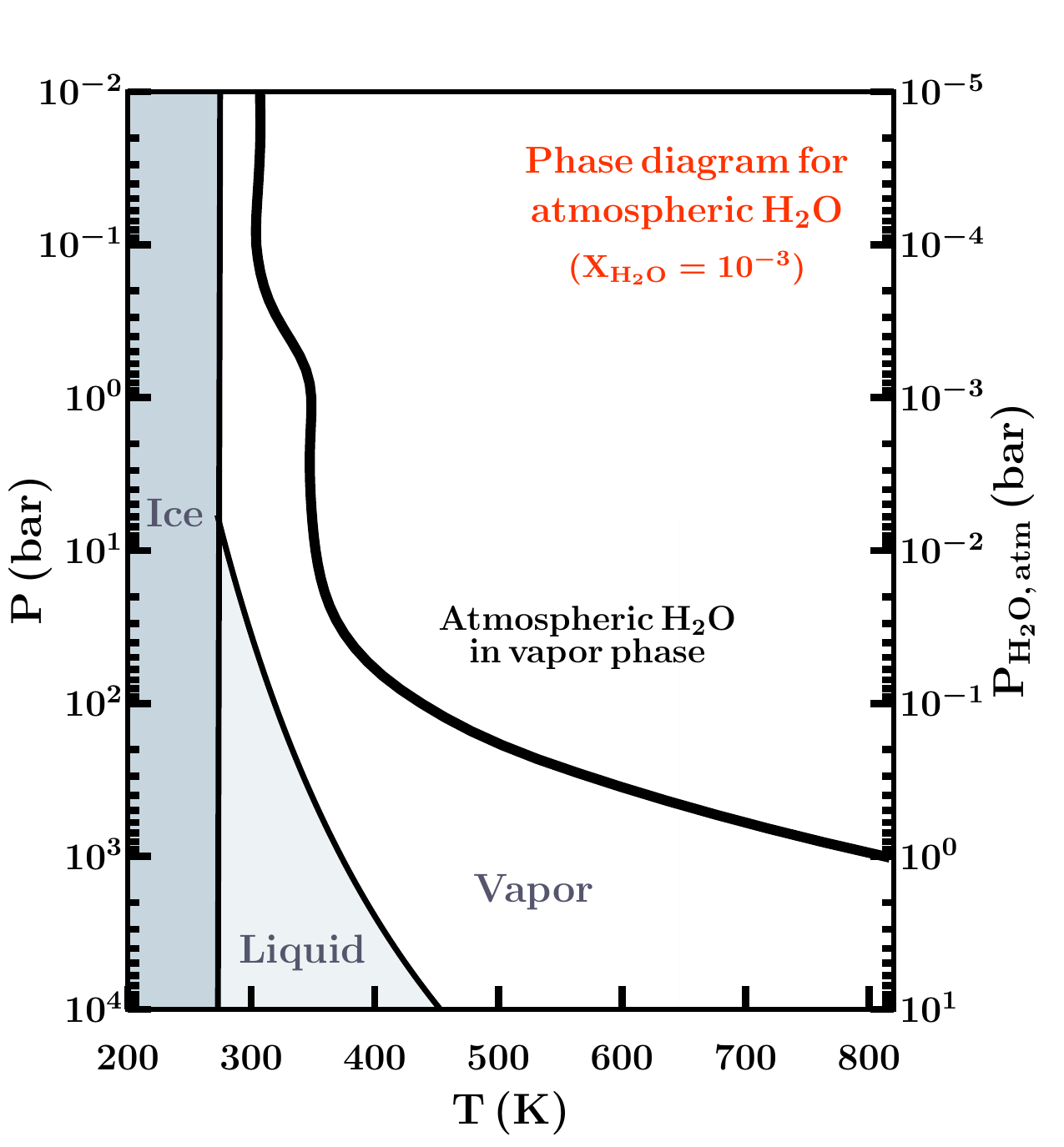}}
        \subfloat{\includegraphics[width=0.48\textwidth]{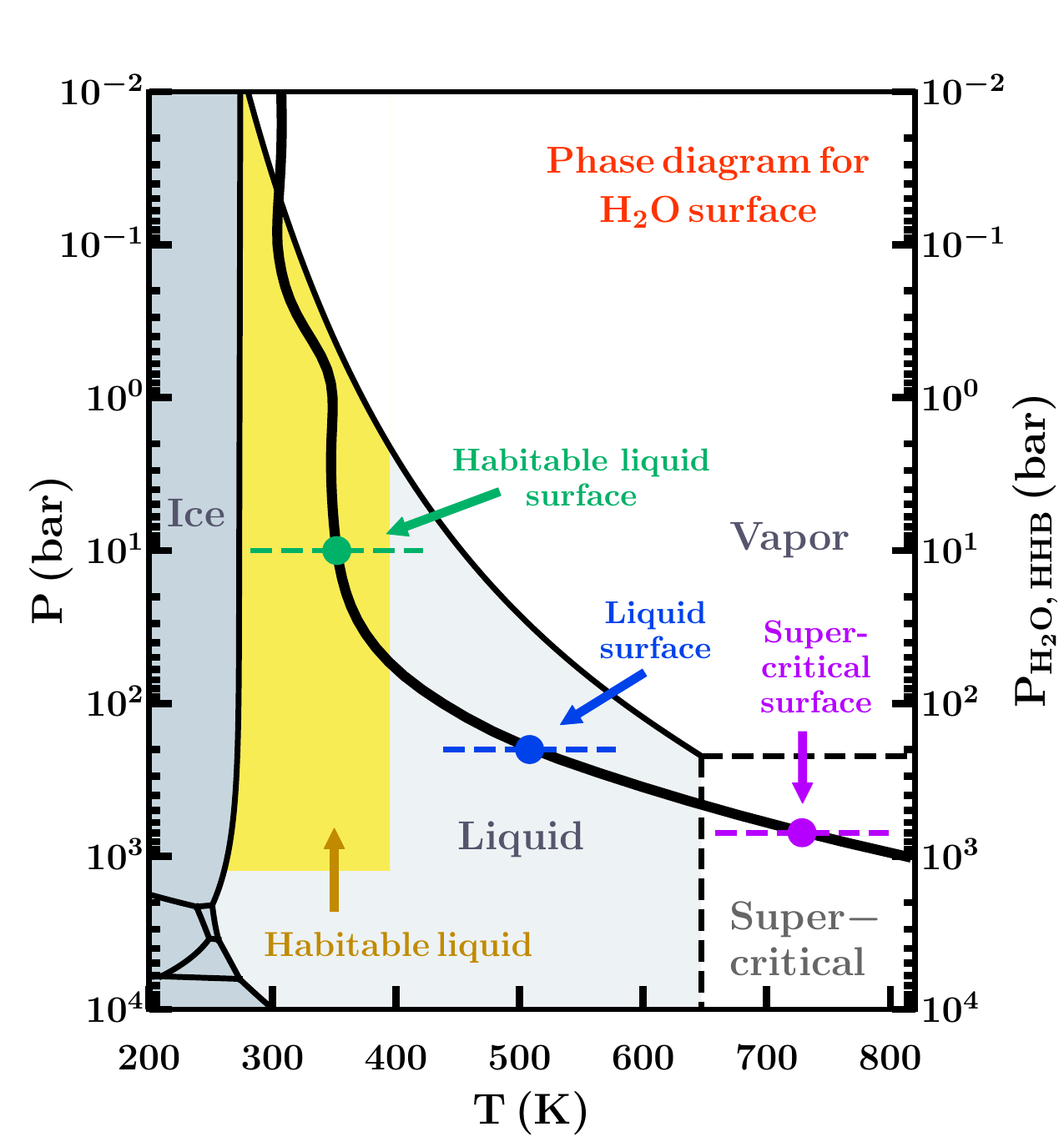}}
    \caption{Schematic of an atmospheric $P$-$T$ profile (bold black lines) and the corresponding phases of the atmospheric H$_2$O (left panel) and of the 100\% H$_2$O layer beneath the atmosphere (right panel). Left panel: background shows the phase diagram for the atmospheric H$_2$O assuming a fixed mixing ratio of $10^{-3}$. For the $P$-$T$ profile shown here, the atmospheric H$_2$O is in the vapor phase throughout. Right panel: background shows the phase diagram for 100\%~H$_2$O, corresponding to the phase at the surface of the 100\%~H$_2$O layer beneath the atmosphere. The region of the liquid phase satisfying $P<1000$~bar, $T<395$~K is highlighted in yellow and corresponds to conditions known to be habitable for extremophiles on Earth \citep[e.g.][]{Merino2019}. Depending on the location of the H$_2$O layer/H$_2$-rich atmosphere boundary (HHB), the surface of the H$_2$O layer in this example can be in the liquid phase (at habitable or inhabitable temperatures) or the supercritical phase. These scenarios are marked by the green, blue and purple dashed lines and circles, respectively. In both panels, the left y-axis ($P$) corresponds to the total atmospheric pressure while the right y-axes ($P_\mathrm{H_2O,\,atm}$ and $P_\mathrm{H_2O,\,HHB}$) show the partial pressure of H$_2$O in the atmosphere (left panel) and the total pressure of H$_2$O at the HHB (right panel), respectively. In the following figures, the H$_2$O phase diagram is always shown for the HHB, as shown in the right panel here.}
    \label{fig:cartoon}
\end{figure*}

\subsection{Thermal Emission Spectra}
\label{sec:spectra}
In this section, we investigate the effects of internal temperature, irradiation, metallicity and clouds/hazes on the thermal emission spectra of mini-Neptunes. These emergent spectra are strongly sensitive to the atmospheric temperature profile, and are therefore impacted by the parameters listed above through their effects on the temperature profile. Figures \ref{fig:spec_clouds} and \ref{fig:spec_main} show thermal emission spectra corresponding to the $P$-$T$ profiles in figures \ref{fig:PT_clouds} and \ref{fig:PT_main}, respectively. In each of these, a flux ratio of 10 ppm is shown by a dashed line to indicate an optimistic minimum uncertainty expected with JWST.

Clouds and hazes predominantly affect the spectrum at shorter wavelengths, as they reflect incident stellar irradiation which peaks in the optical. However, this reflected light is well below an optimistic 10 ppm uncertainty for JWST, and is, therefore, unlikely to be detectable for the temperate mini-Neptunes modelled here. Nevertheless, clouds and hazes can also affect the spectrum through their effects on the $P$-$T$ profile. For example, high-altitude ice clouds significantly cool the photospheric temperature, which results in less emitted flux. In the case of hazes and deeper KCl clouds, stronger haze/cloud opacity results in a more isothermal $P$-$T$ profile in the photosphere and weaker absorption features. This can be seen in the continuum peak at $\sim$5~$\mu$m, which is smaller for models with stronger haze/KCl cloud opacity.

Although $T_\mathrm{irr}$ and $T_\mathrm{int}$ both play major roles in determining the $P$-$T$ profile and radiative-convective boundary (section \ref{sec:TeqTint}), only $T_\mathrm{irr}$ has a significant effect on the observable spectrum. As $T_\mathrm{irr}$ translates the $P$-$T$ profile to hotter/cooler temperatures, the emergent flux increases/decreases accordingly. However, $T_\mathrm{int}$ largely affects the $P$-$T$ profile below the photosphere and does not impact the spectrum unless its value is sufficiently high. For the models shown here with $T_\mathrm{irr}$=350~K, $T_\mathrm{int}$=200~K is enough to affect the photosphere and result in a slightly higher continuum peak at $\sim$5 $\mu$m. This is a stronger effect for hazy models compared to clear ones (left column of figure \ref{fig:spec_main}), though the haze also mutes the continuum feature. Estimates of $T_\mathrm{int}$ for mini-Neptunes will therefore rely on theoretical cooling models as this parameter will not be derivable from observed spectra unless it is very high.

The right column of figure \ref{fig:PT_main} shows that metallicity does not have a strong effect on the observable spectrum, especially at longer wavelengths. As discussed by \citet{Spiegel2010}, this is because the increase in temperature due to increased metallicity is balanced by the photosphere shifting to lower pressures. Furthermore, this shallower photosphere is also more isothermal, which weakens the strength of spectral features.

While each atmospheric parameter affects the observable spectrum to different extents, we note that all of the spectra in figures \ref{fig:spec_clouds} and \ref{fig:spec_main} are well above 10~ppm at longer wavelengths and should be observable with JWST. In section \ref{sec:observability}, we discuss how JWST observations could help to constrain the conditions in atmospheres of mini-Neptunes.

\section{Considerations for habitability}
\label{sec:case_studies}

In the search for extrasolar habitability, planets in the habitable zones of M-dwarfs provide excellent targets thanks to their small host star radii and semi-major axes. Furthermore, mini-Neptunes have large atmospheric scale heights and planet/star size ratios, making them especially conducive to atmospheric characterisation. K2-18~b is a prime example of such a planet, and its atmosphere has been characterised through transmission spectroscopy, leading to a strong detection of water vapour \citep{Benneke2019b,Tsiaras2019}. Using the observed spectrum and bulk properties (mass and radius), \citet{Madhusudhan2020} placed joint constraints on the atmosphere and interior of K2-18b and found several solutions allowing for liquid water at habitable temperatures and pressures at its surface. The Transiting Exoplanet Survey Satellite (TESS) is expected to find more planets of this type, with several existing candidates and some already confirmed \citep[e.g.][]{Gunther2019}. 

Traditional definitions of the habitable zone are typically designed for terrestrial planets with thin atmospheres \citep{Kasting1993,Kopparapu2017,Meadows2018book}, for which the equilibrium temperature can be comparable to the surface temperature. This is not the case for mini-Neptunes, which host H$_2$-rich envelopes and whose surfaces can occur at much deeper/higher pressures. Due to the greenhouse effect of H$_2$, the temperatures at these higher pressures can be significantly hotter than the equilibrium temperature of the planet \citep{Pierrehumbert2011,Koll2019}. Habitability on mini-Neptunes therefore relies on physical processes which can counter this heating effect, such as clouds and hazes. 

In this section, we explore atmospheric conditions under which K2-18~b could host liquid water on its surface at habitable temperatures and pressures. Liquid water on the surface is typically considered to be a nominal requirement for habitability \citep[e.g.][]{Meadows2018book}. Furthermore, it is known that extremophiles on Earth are able to survive at temperatures up to 395~K and pressures up to 1250~bar \citep{Merino2019}. In what follows, we therefore refer to temperatures $\lesssim$395~K at pressures $\lesssim$1250~bar as habitable conditions. We begin by discussing the requirements for a liquid ocean in section \ref{sec:ocean_conditions}. Since the atmospheric H$_2$O abundance above an ocean can be a complex function of altitude (as seen on Earth, e.g., \citealt{Pierrehumbert2006}), we then explore atmospheric $P$-$T$ profiles for two end-member scenarios: fixed metallicity in section \ref{sec:hab_subsat} and 100\% relative humidity in section \ref{sec:hab_sat}.

\begin{figure*}
    \centering
        \subfloat{\includegraphics[width=0.48\textwidth]{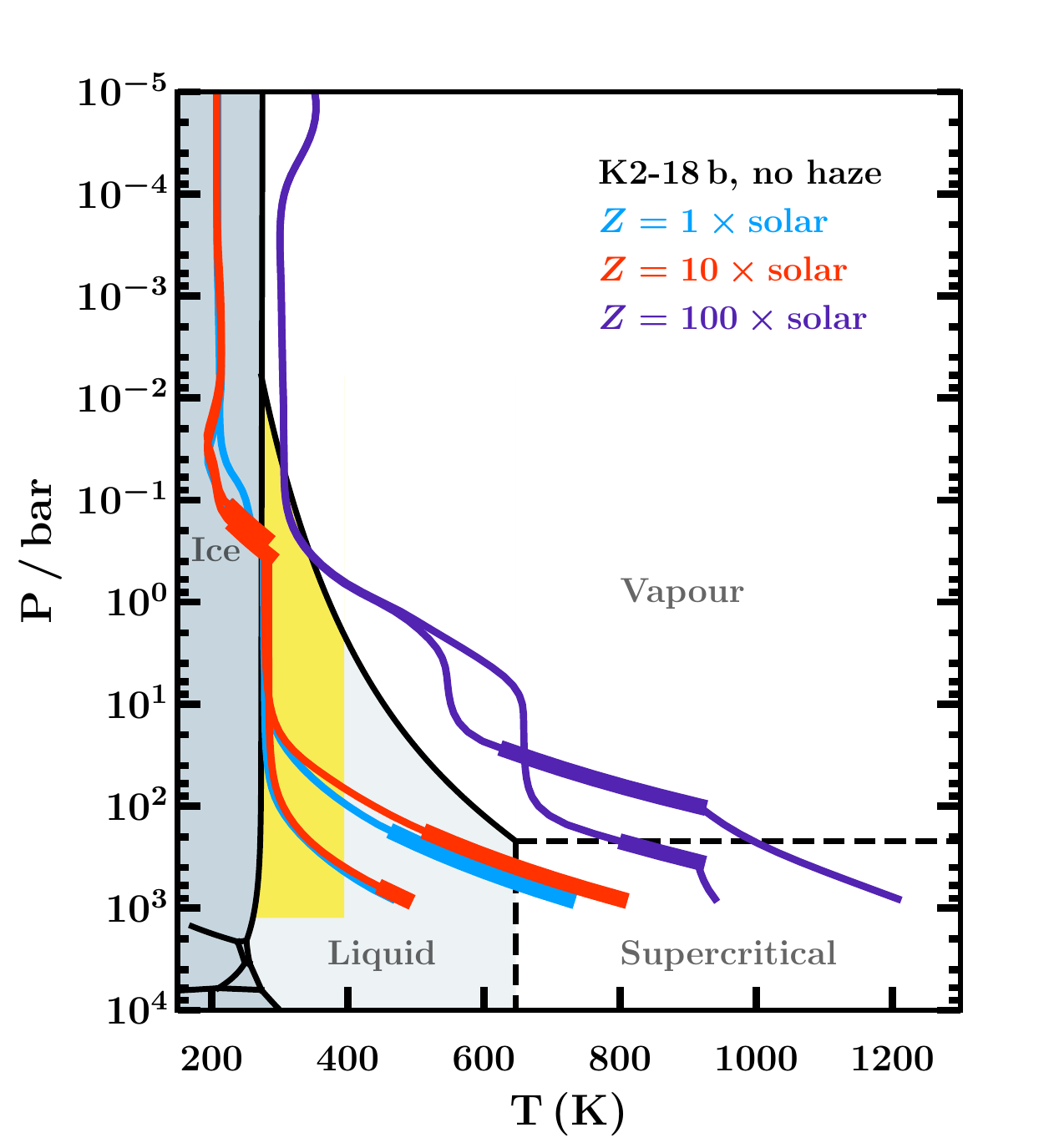}}
        \subfloat{\includegraphics[width=0.48\textwidth]{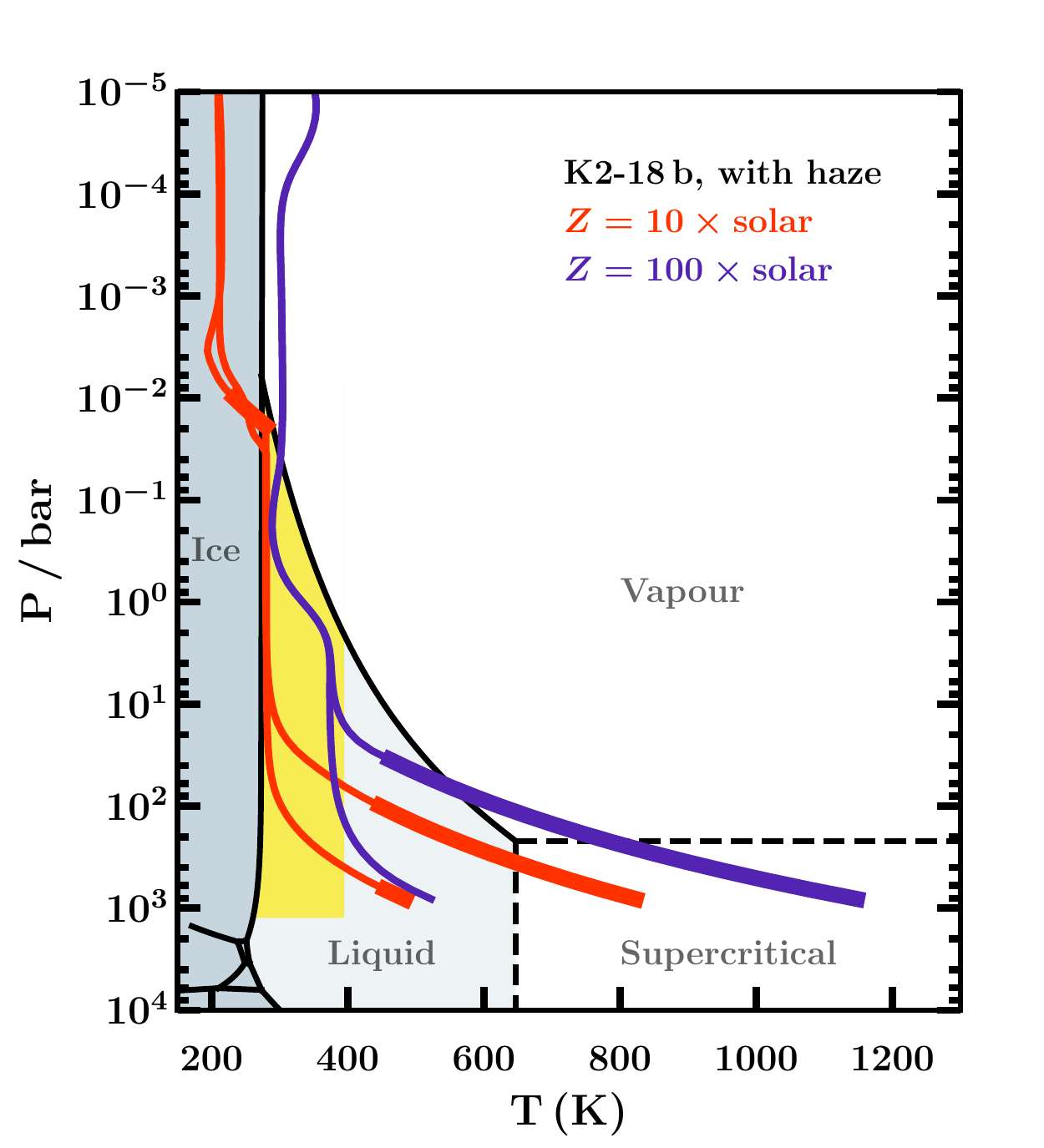}}
    \caption{Model dayside $P$-$T$ profiles for K2-18~b, for a range of metallicities and internal temperatures, that could allow habitable conditions below the atmosphere. For each metallicity, the cooler/hotter profile at 1000 bar corresponds to $T_\mathrm{int}$=25~K/50~K. Left panel: models with no haze. Models with 1$\times$ and 10$\times$ solar metallicity enter the water ice phase so we include ice clouds where they are thermodynamically required (cloud bases in the range 0.2-0.3~bar). The 100$\times$solar metallicity model has KCl and ZnS clouds in the deep atmosphere (cloud bases between 100 and 1000~bar). Right panel: models with hazes. The 10$\times$solar metallicity model with $T_\mathrm{int}$=25 K has haze equivalent to 500$\times$ H$_2$ Rayleigh scattering, while the other models have 1000$\times$ H$_2$ Rayleigh scattering. Models with 10$\times$solar metallicity have water ice clouds (cloud bases in the range 0.01-0.03~bar). The phase diagram of H$_2$O is shown in the background, corresponding to the surface conditions of a 100\% H$_2$O layer beneath the H$_2$-rich atmosphere, i.e., at the HHB (see section ~\ref{sec:case_studies} and right panel of figure \ref{fig:cartoon}). The shaded yellow regions show habitable conditions in Earth's oceans, with T$<$395~K and P$<$1250~bar \citep{Merino2019}. Models passing through the yellow region could therefore potentially host habitable liquid water at the HHB.}
    \label{fig:300}
\end{figure*}

\begin{figure*}
    \centering
        \subfloat{\includegraphics[width=0.48\textwidth]{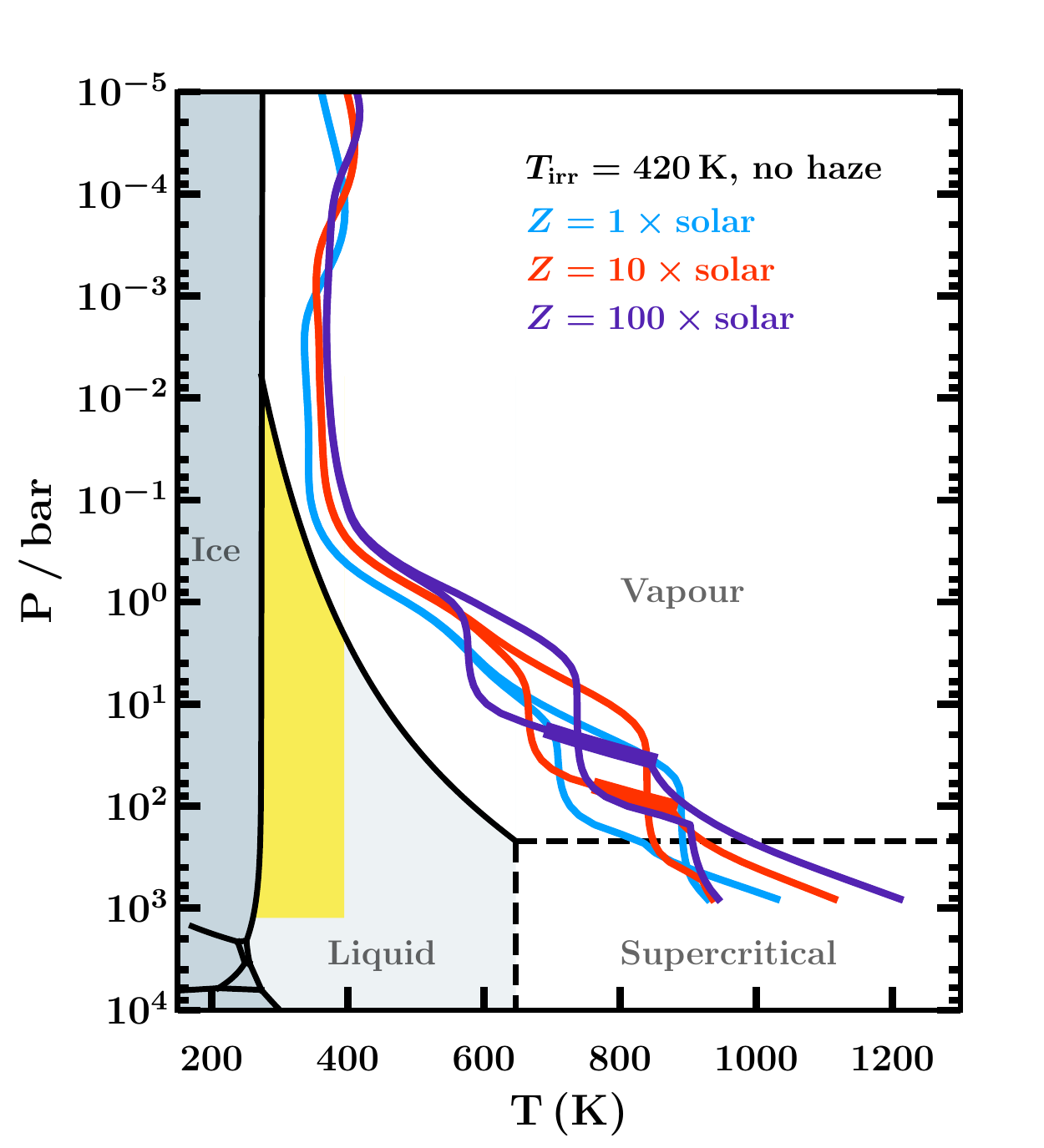}}
        \subfloat{\includegraphics[width=0.48\textwidth]{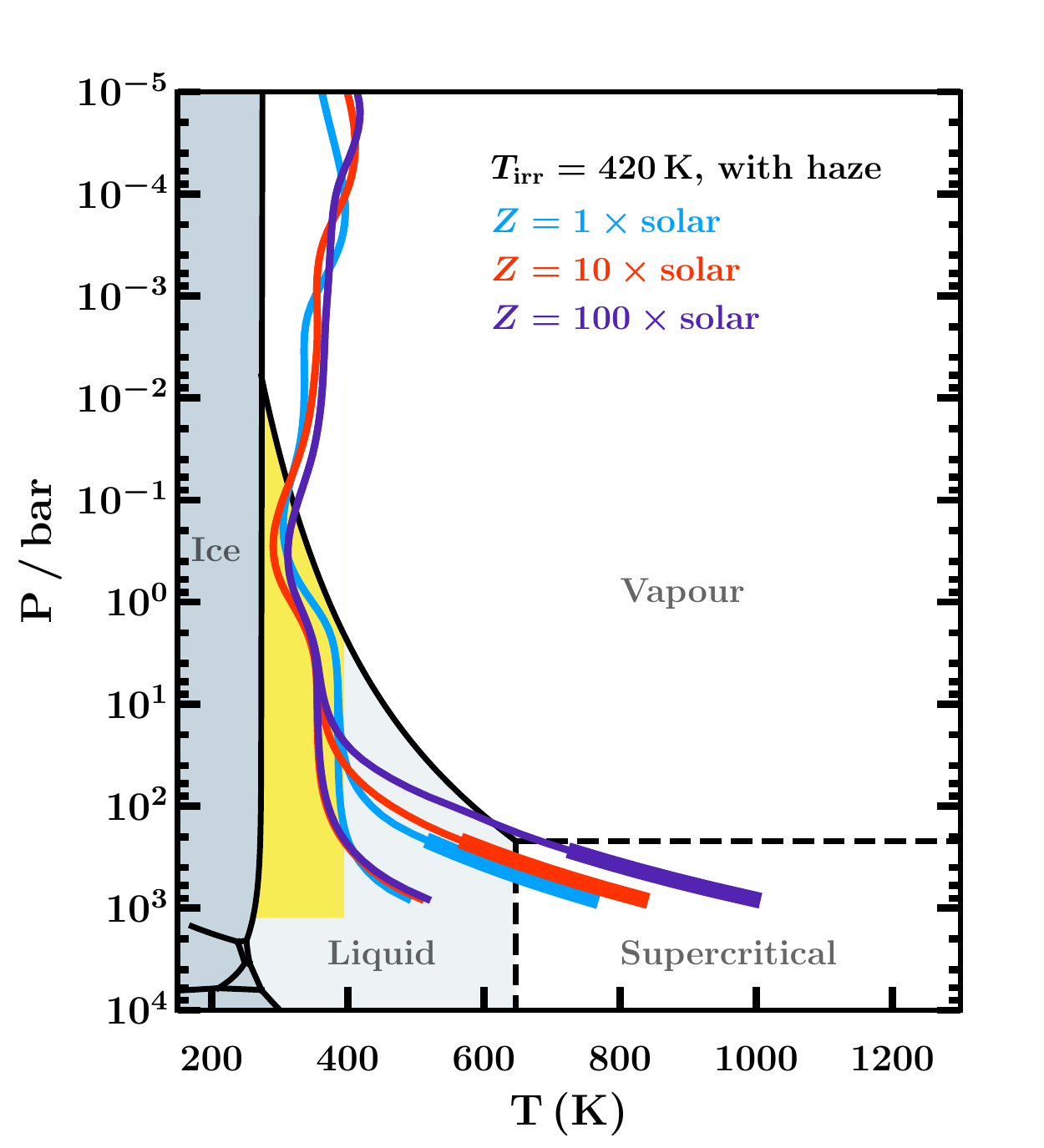}}
    \caption{Same as figure \ref{fig:300}, but for a planet resembling K2-18~b with an irradiation temperature of 420~K (i.e. receiving $\sim$2.5$\times$ more incident flux). Left panel: models with no hazes. Each model includes KCl and ZnS clouds with cloud base pressures between 30 and 1000 bar. Right panel: models including hazes. 1$\times$solar models include haze equivalent to 3000$\times$ H$_2$ Rayleigh scattering, while the other models have 10000$\times$ H$_2$ Rayleigh scattering. Strong haze opacity is required for the $P$-$T$ profiles to enter the shaded yellow region, where a H$_2$O surface beneath the atmosphere could host habitable liquid water.}
    \label{fig:400}
\end{figure*}

\begin{figure*}
    \centering
    \subfloat{\includegraphics[width=0.49\textwidth]{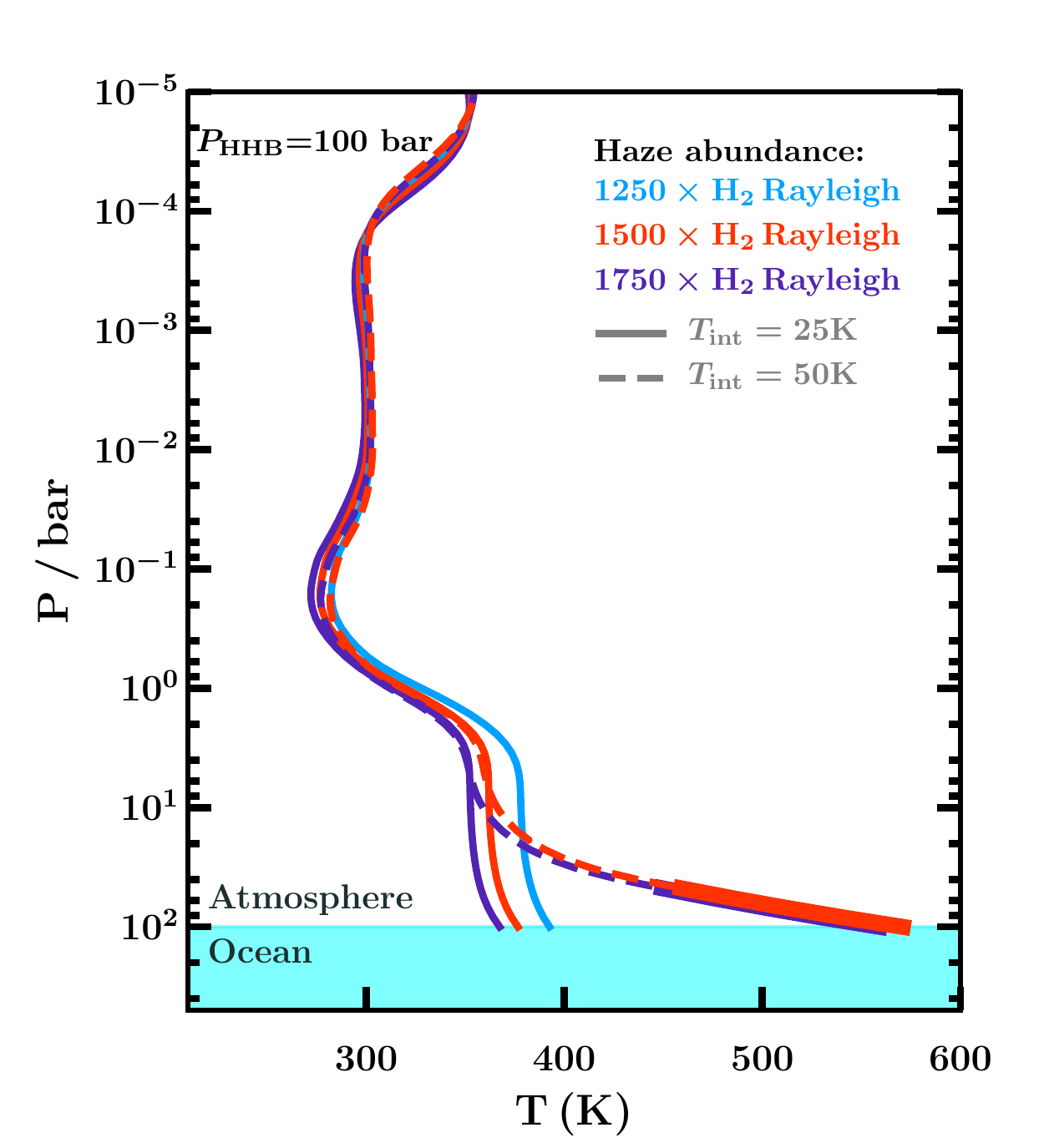}}
    \subfloat{\includegraphics[width=0.49\textwidth]{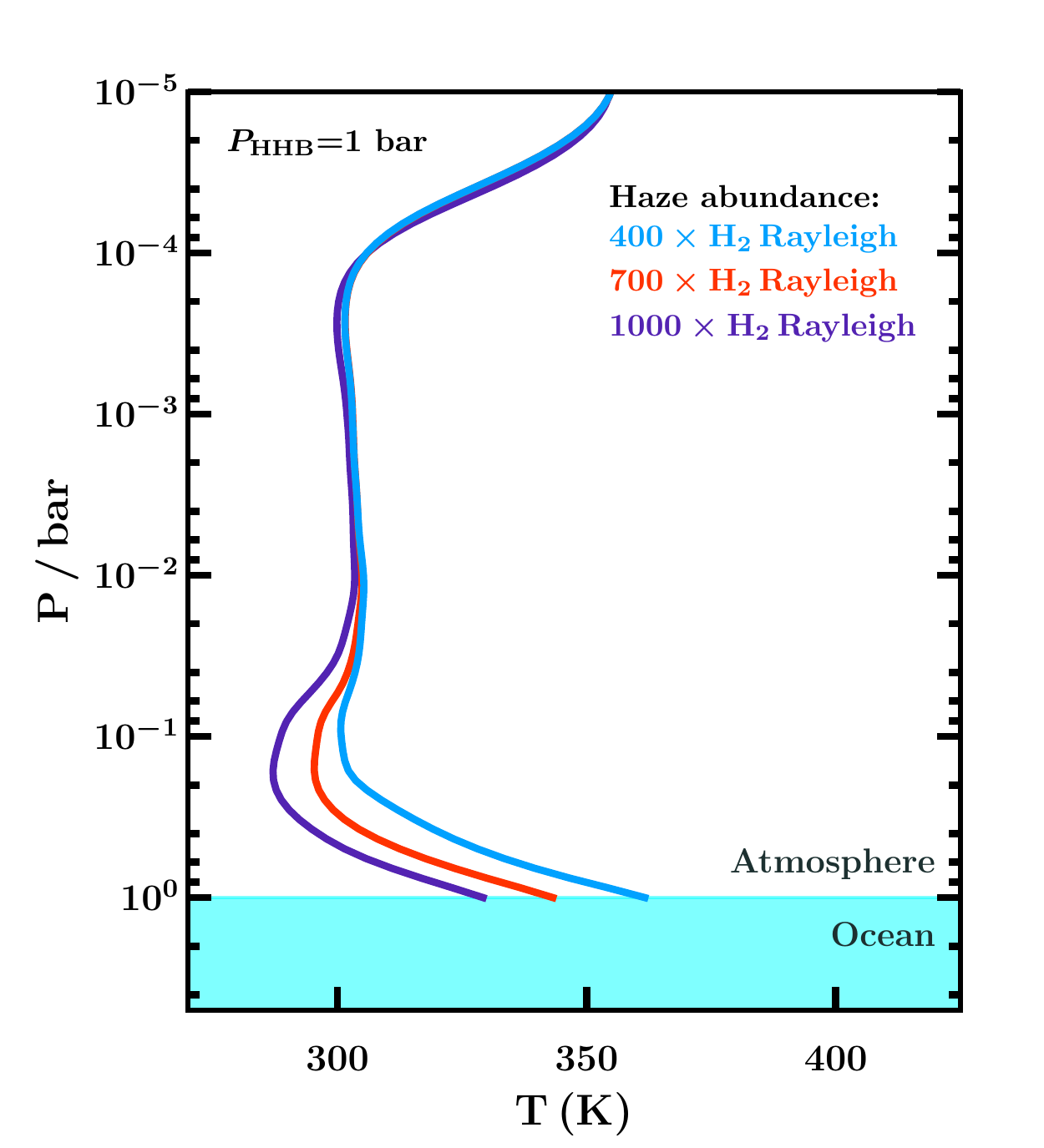}}
    \caption{Model dayside $P$-$T$ profiles for K2-18~b with 100\% relative humidity which allow for a liquid ocean at the surface. Surface temperatures below $\sim$400~K could allow for habitable conditions for Earth-like aquatic life \citep[e.g.][]{Merino2019}. Left panel: Models with an ocean-atmosphere boundary (HHB) at 100~bar and varying haze abundances and internal temperatures (see legend). Right panel: Models with the HHB at 1~bar and varying haze abundances; these $P$-$T$ profiles are not sensitive to $T_\mathrm{int}$ between 25-50~K. The atmospheric H$_2$O abundance is fully saturated between the ocean surface and the cold trap (see section~\ref{sec:hab_sat}).} 
    \label{fig:sat_PTs}
\end{figure*}

\subsection{Conditions for an ocean}
\label{sec:ocean_conditions}
An essential condition for the presence of a water ocean in a planet is the availability of a water reservoir beneath the atmosphere at the right thermodynamic conditions. Several works have explored the possibility of water oceans in super-Earths and water worlds with a large H$_2$O layer in the interior \citep[e.g.,][]{Leger2004,Rogers2010b,Rogers2011,Zeng2013,thomas2016} as well as ice giants with mixed-composition interiors that require sufficiently large H$_2$O mixing ratios at low photospheric temperatures \citep[e.g.,][]{Wiktorowicz2007}. In the present work, we assume the presence of a water layer below the H$_2$-rich atmosphere, following the recent constraints on the mini-Neptune K2-18b from \citet{Madhusudhan2020}. Whether or not a liquid water ocean is possible below the atmosphere, thus, depends on the pressure-temperature conditions at the bottom of the atmosphere as we model here. 

For K2-18b, \citet{Madhusudhan2020} find that the atmospheric mass fraction is between $\lesssim 10^{-6}$ and $\sim 6\times 10^{-2}$, allowing surface pressures as low as $\sim$1~bar for some atmospheric $P$-$T$ profiles. In their interior model, a 100\% H$_2$O layer exists below the atmosphere, similar to other models in the literature for super-Earths and mini-Neptunes \citep[e.g.][]{Nettelmann2011,Rogers2011,Valencia2013}. As discussed in section \ref{sec:intro}, not all mini-Neptunes would have masses and radii consistent with habitable pressures ($\lesssim$1000~bar) below their H$_2$-rich atmospheres. However, mini-Neptunes with bulk densities similar to or higher than K2-18~b could allow for habitable surface pressures. 

In what follows, we use the term HHB, as in \citet{Madhusudhan2020}, to denote the boundary between the H$_2$O surface and the H$_2$-rich atmosphere, i.e., the pressure level where the atmosphere terminates and the 100\% H$_2$O layer begins. Just above the HHB, the H$_2$O abundance is that of the atmosphere, i.e. lower than 100\%. At the HHB and below, the H$_2$O abundance is 100\%, i.e., in the water layer. Therefore, the phase of water can change across the HHB depending on the pressure and temperature conditions at the HHB; for example, from vapour in the atmosphere to liquid in the 100\% H$_2$O layer below. To assess the phase of the 100\% H$_2$O layer just below the HHB, we compare model atmospheric $P$-$T$ profiles to the phase diagram of 100\% H$_2$O \citep{fei1993,wagner2002,seager2007,french2009,sugimura2010,thomas2016}, as shown in the right panel of figure \ref{fig:cartoon}. Note that this phase diagram does not correspond to the atmospheric H$_2$O, as the partial pressure of H$_2$O in the atmosphere is less than the total pressure (see left panel of figure \ref{fig:cartoon}). Instead, at each pressure and temperature it corresponds to the phase of H$_2$O at the surface of the 100\% H$_2$O layer below assuming that the bottom of the atmosphere is at that pressure and temperature.

\subsection{Effects of cloud/hazes, metallicity and $T_\mathrm{int}$}
\label{sec:hab_subsat}

In this section, we explore the effects of clouds, hazes, metallicity and internal temperature on the potential habitability of K2-18~b. For these models, we assume a fixed atmospheric metallicity informed by the transmission spectrum of K2-18~b, as described below. The assumed H$_2$O abundance is largely sub-saturated in these models, representing one extreme in relative humidity, which is known to be a complex function of many parameters on Earth \citep[e.g.][]{Pierrehumbert2006}. We then explore models with a saturated H$_2$O abundance in section \ref{sec:hab_sat}. 

Our model $P$-$T$ profiles are shown in figures~\ref{fig:300} and ~\ref{fig:400} and discussed further below. At any given pressure, $P$, in the $P$-$T$ profile, the comparison to the H$_2$O phase diagram indicates the phase at the surface of the H$_2$O layer if the HHB occurred at that pressure (i.e., $P_{\rm HHB}=P$), as shown in the right panel of figure \ref{fig:cartoon}. Note that we do not assume a particular location for the HHB in our models. Instead, we calculate the atmospheric $P$-$T$ profile up to 1000~bar, and see where the HBB would cross the H$_2$O phase diagram for different values of $P_{\rm HHB}$. For example, if a model $P$-$T$ profile crosses the liquid phase for 100\% H$_2$O at 10~bar, this suggests that if the HHB were to occur at 10~bar, the surface of the 100\% H$_2$O layer would be liquid. 

\citet{Madhusudhan2020} show that $P_{\rm HHB}$ $\gtrsim$ 1~bar for K2-18~b. That is, the surface between the atmosphere and the 100\% H$_2$O layer can exist at pressures $\gtrsim$ 1~bar. Therefore, atmospheric $P$-$T$ profiles which intersect with the liquid H$_2$O phase at pressures $\gtrsim 1$ bar could allow for surface liquid water below the atmosphere at that pressure. Furthermore, if the $P$-$T$ profile intersects the liquid phase at $<395$~K, this liquid surface water would be at habitable temperatures. We reiterate that the phase diagram of H$_2$O shown here does not correspond to the phase of H$_2$O in the atmosphere. Instead, it corresponds to the phase of the 100\% H$_2$O layer at the HHB, just below the atmosphere.

In order to generate atmospheric $P$-$T$ profiles for K2-18~b, we use spectroscopic constraints on the atmospheric chemistry from \citet{Madhusudhan2020}. \citet{Madhusudhan2020} find an atmospheric H$_2$O abundance of $\sim$1-100$\times$ solar, with no detection of any other species. We therefore include only H$_2$O in our models, varying its metallicity from 1 to 100$\times$ solar. In all our models, we also include salt (KCl/ZnS) clouds and water ice clouds wherever they are thermochemically expected to occur (section \ref{sec:model}). Water ice cloud particle sizes can have a wide range of modal particle sizes \cite[e.g.][]{Morley2014}; we tried a range of sizes and in the models presented here use particle sizes in the range 1-3 $\mu$m. The base of each ice cloud is located approximately where the atmospheric H$_2$O enters the ice phase (as described in section \ref{sec:opacity}). In the models shown in figure \ref{fig:300}, the cloud bases lie in the range 0.01-0.2~bar. As described in section \ref{sec:model}, we calculate models for two end-member values of $T_\mathrm{int}$: 25 and 50~K.

We find that a range of atmospheric conditions allow for habitability in K2-18~b. \citet{Madhusudhan2020} find habitable solutions for two atmospheric $P$-$T$ profiles. Here, we explore a wider range of parameters and their impact on potential habitability. Figure \ref{fig:300} shows model $P$-$T$ profiles for K2-18~b both with and without hazes (right and left panels, respectively). As expected, higher-metallicity models are hotter and host salt clouds in the lower atmosphere, while the low-metallicity models are cooler. The haze-free models with 1 and 10$\times$solar metallicity are sufficiently cool that ice clouds are present in the upper atmosphere, which further cools the deeper regions of the atmosphere (light blue and red lines in left panel of figure \ref{fig:300}, respectively). Higher-metallicity models with hazes also remain cool at high pressures (right panel of figure \ref{fig:300}). For both the hazy models and the haze-free 1 and 10$\times$ solar models, a H$_2$O surface beneath the atmosphere could potentially host habitable liquid surface water if this surface occurs at the right pressure (see shaded yellow regions in figure~\ref{fig:300}).

We find that a planet resembling K2-18~b but with a higher irradiation temperature can also be potentially habitable. Figure \ref{fig:400} shows model $P$-$T$ profiles for a planet like K2-18~b but with $T_\mathrm{irr}$=420~K rather than 332~K (i.e., receiving $\sim$2.5$\times$ more incident flux, achieved by decreasing the semi-major axis). With no hazes or ice clouds (left panel), these profiles reach high temperatures deep in the atmosphere. In this case, the models do not cross the `potentially habitable' shaded yellow region. However, models with strong haze opacity (right panel) are significantly cooled such that habitable liquid water could exist at the surface of a H$_2$O layer beneath the atmosphere. Therefore, mini-Neptunes similar to but somewhat warmer than K2-18~b could also have the potential for habitability, although stronger optical opacity is needed to cool such planets sufficiently. We also note that throughout this work we have assumed no day-night redistribution of the incident radiation. Therefore, our habitability estimates are conservative. Allowing for efficient day-night redistribution could allow such planets at even higher $T_\mathrm{irr}$ to be potentially habitable. We discuss this further in section~\ref{sec:discussion}.

\subsection{Effects of H$_2$O saturation}
\label{sec:hab_sat}

In this section, we explore models for K2-18~b for which the atmosphere is saturated with H$_2$O vapour, representing the upper extreme in relative humidity. We consider two surface pressures corresponding to the base of the atmosphere: $P_\mathrm{HHB}$=100~bar (left panel of figure \ref{fig:sat_PTs}) and $P_\mathrm{HHB}$=1~bar (right panel of figure \ref{fig:sat_PTs}). For each surface pressure, we consider a range of haze opacities and $T_\mathrm{int}$=25-50~K to explore how these parameters affect the habitability of the ocean. 

We assume the atmosphere to be 100\% saturated between the ocean surface and the cold trap at higher altitudes. The mixing ratio of water vapor follows the saturation curve up to the cold trap where it reaches the minimum value in saturation for the corresponding temperature profile. The pressure level at this minimum is $\sim$0.1-0.3~bar for most cases, but can be as deep as 20 bar for cases with P$_{\rm HHB}$=100~bar and T$_{int}$=50~K. Beyond the cold trap the H$_2$O saturation vapour pressure begins to increase with altitude and the actual H$_2$O mixing ratio there depends on various factors including the efficiency of atmospheric mixing \citep[e.g.][]{Pierrehumbert2006}. At altitudes beyond the cold trap (i.e., at lower pressures) we assume the H$_2$O mixing ratio to be constant at either the minimum value at the cold trap or 100$\times$solar, whichever is smaller. The maximum abundance of 100$\times$solar is set by the upper-limit on the H$_2$O abundance at the photosphere ($\sim$1-100 mbar level) derived from the observed transmission spectrum of K2-18~b \citep{Madhusudhan2020}.

We note that in some cases, especially when the temperature structure is nearly isothermal in the lower atmosphere, a second cold trap can occur close to the ocean surface. Here, the saturation vapour pressure increases with altitude starting right at the surface of the ocean, before decreasing again higher up in the atmosphere and leading to the more conventional cold trap there as discussed above. For such ``surface cold traps" we assume that atmospheric mixing easily overcomes the cold trap and  causes the H2O abundance to remain saturated in the lower atmosphere.

In figure \ref{fig:sat_PTs}, we show models for which the surface of the 100\% H$_2$O layer is in the liquid phase. For a surface pressure of 100~bar (left panel of figure \ref{fig:sat_PTs}), we find several cases which allow for habitable temperatures at the ocean surface (i.e. $\lesssim$400~K). In particular, models with $T_\mathrm{int}$=25~K and haze abundances above $\sim$1250$\times$H$_2$ Rayleigh scattering satisfy this condition. Models with higher $T_\mathrm{int}$ and/or less haze have warmer surface temperatures but still allow for a liquid ocean. We note that since the definition of habitability used here is based solely on Earth-like life, these warmer temperatures do not preclude unknown life forms which may have adapted to more extreme conditions.

For a surface pressure of 1~bar (right panel of figure \ref{fig:sat_PTs}), we find that even more solutions allow for habitable conditions at the ocean surface. Firstly, at these low pressures, a $T_\mathrm{int}$ in the range 25-50~K does not affect the $P$-$T$ profile, allowing for more habitable cases. Secondly, much less haze is needed to achieve a habitable surface temperature ($\gtrsim$~400$\times$H$_2$ Rayleigh scattering) as the atmosphere is thinner and has less greenhouse warming. We therefore find that despite the high relative humidity (i.e., 100\% saturation) above the ocean, a range of surface pressures, haze opacities and $T_\mathrm{int}$ allow for potentially habitable conditions in planets like K2-18~b. 

\label{sec:obs_miri_hab}
\begin{figure*}
    \centering
    \includegraphics[width=\textwidth]{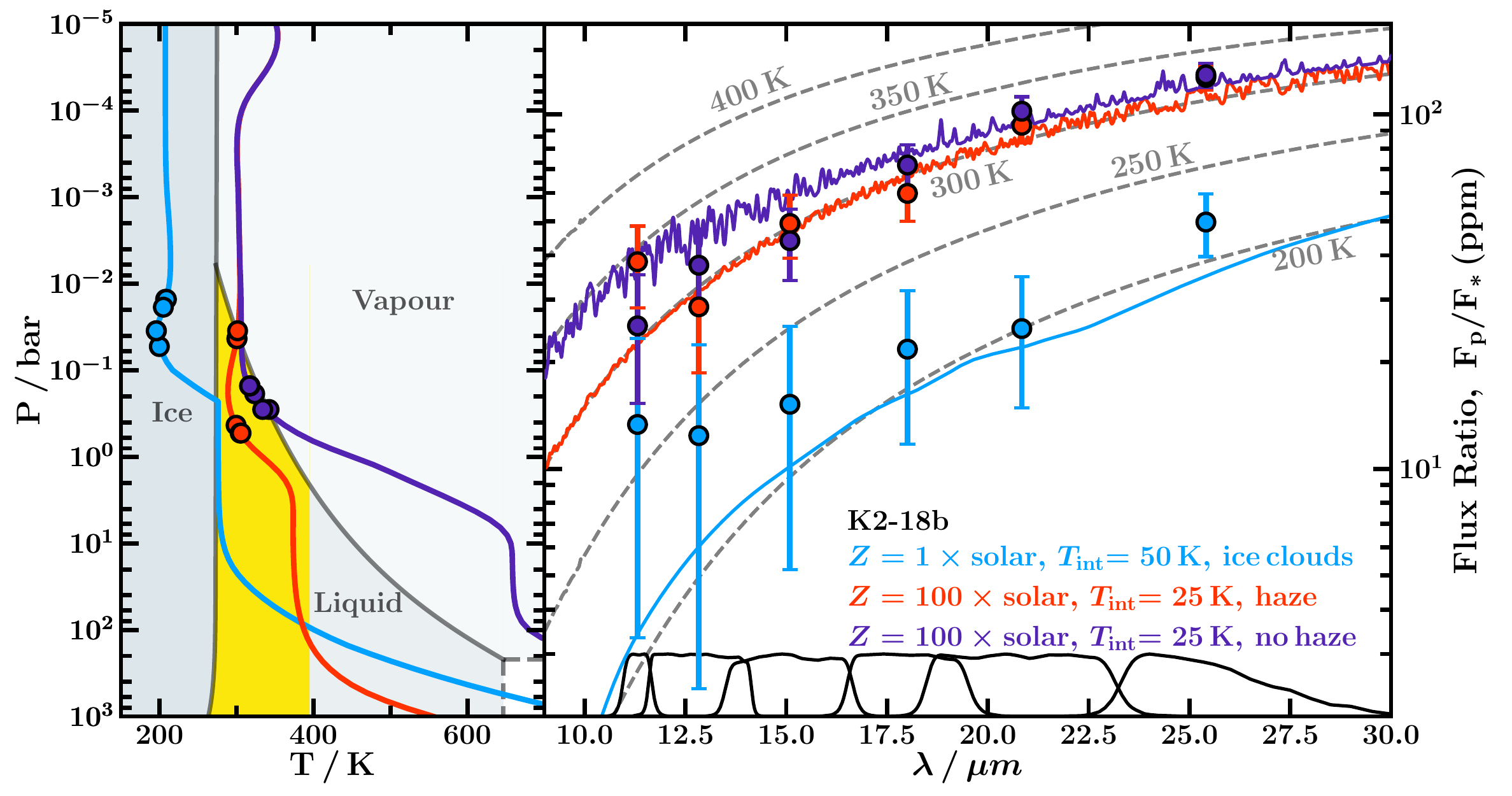}
    \caption{Observability of temperate and potentially habitable mini-Neptunes with JWST/MIRI. Left: Model $P$-$T$ profiles for K2-18~b with varying metallicities (1 and 100$\times$solar) and internal temperatures ($T_\mathrm{int}$=25 and 50 K). Clouds are included wherever they are thermodynamically expected (see figure \ref{fig:300} for details). Phase diagram for 100\% H$_2$O shows the phase that a water layer would have at the boundary between the H$_2$-rich atmosphere and H$_2$O interior (HHB). The yellow region highlights pressures and temperatures where this water would be liquid and habitable. The coloured circles denote brightness temperatures in different JWST/MIRI photometric bands described in the right panel. Right: Model spectra corresponding to the $P$-$T$ profiles in the left panel. Simulated photometric data from the JWST/MIRI filters at wavelengths $>$10 $\mu$m are shown for each spectrum (coloured circles, see section \ref{sec:observability}), with best-case error bars of 10~ppm. The brightness temperatures for the central unperturbed value of each photometric point are plotted on the corresponding $P$-$T$ profile in the left panel, showing the depths probed by each measurement. Sensitivity curves for each filter are shown in black. Grey dashed lines show blackbody curves corresponding to planetary temperatures of 200, 250, 300, 350 and 400 K.  }
    \label{fig:hab_obs}
\end{figure*}

\begin{figure*}
    \centering
    \includegraphics[width=\textwidth]{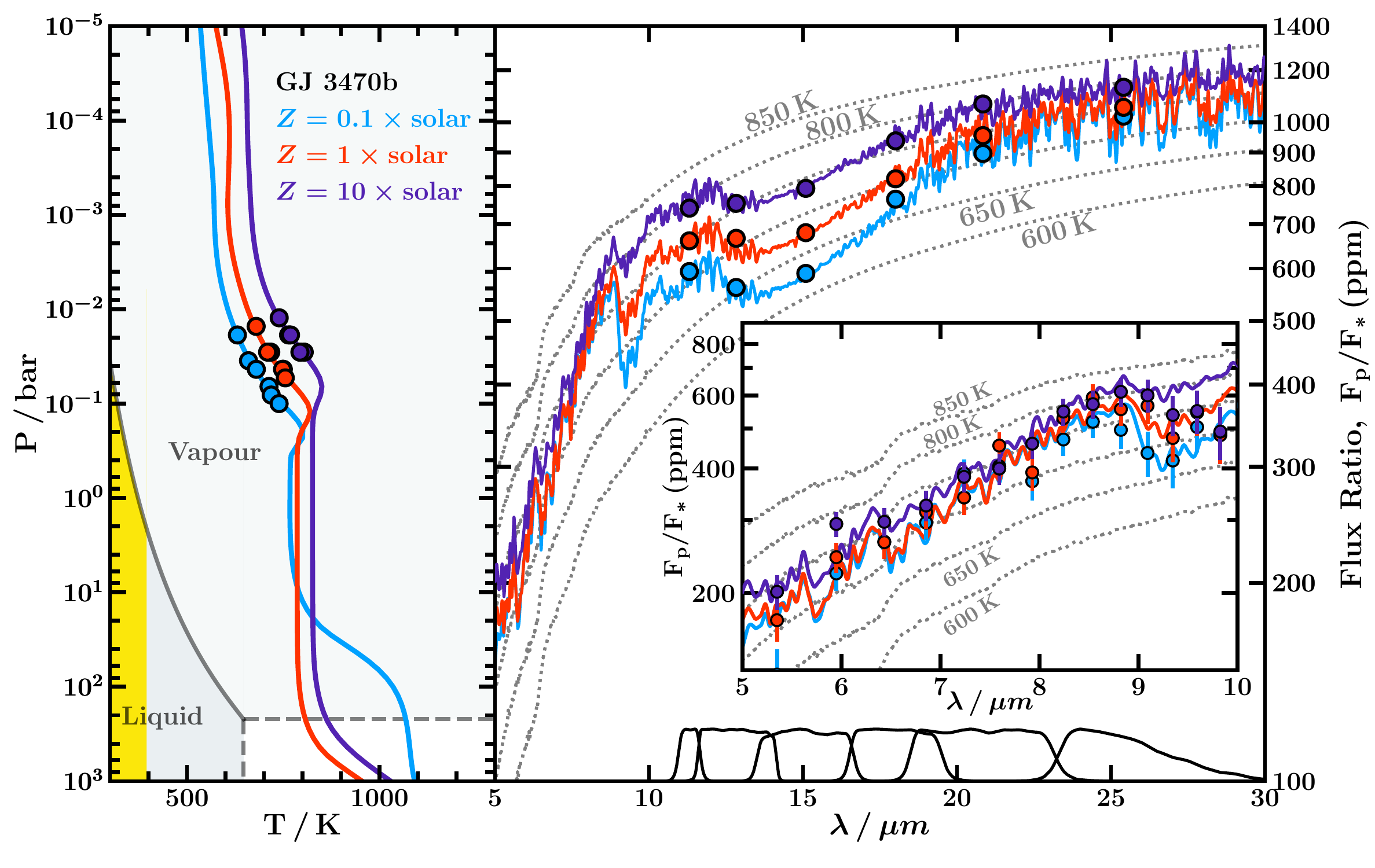}
    \caption{Observability of thermal emission from warm mini-Neptunes with JWST/MIRI. Left panel: model $P$-$T$ profiles for GJ~3470~b with varying metallicities (0.1, 1 and 10 $\times$solar) and $T_\mathrm{int}$=30 K (see section \ref{sec:obs_miri_steam}). Phase diagram shows the phase of a 100\% H$_2$O layer at the HHB, as in figure \ref{fig:hab_obs}. The coloured circles denote brightness temperatures in different JWST/MIRI photometric bands described in the right panel. Right panel: Spectra corresponding to these $P$-$T$ profiles, with simulated MIRI photometry data (coloured circles, see section \ref{sec:observability}). Brightness temperatures for the central value of each photometric point are plotted on the corresponding $P$-$T$ profile in the left panel. We assume error bars of 10~ppm, which are smaller than the data symbols. Sensitivity curves for each MIRI filter used are shown in black, as in figure \ref{fig:hab_obs}. Inset: Enlarged view of spectra in the range 5-10~$\mu$m. Simulated MIRI~LRS data, generated using PandExo \citep{Pandexo}, are shown as coloured circles and error bars. Dashed grey lines show blackbody curves for planetary temperatures of 600, 650, 700, 750, 800 and 850 K.}
    \label{fig:steamy_obs}
\end{figure*}

\section{Observability}
\label{sec:observability}
As we have shown above, mini-Neptunes such as K2-18~b could provide potential targets in the search for habitability in exoplanets. In particular, their observability with current and future facilities arguably makes them optimal targets for such studies. Recent observations of K2-18~b \citep{Benneke2019b,Tsiaras2019} have already shown that current facilities are able to observe temperate mini-Neptunes with transmission spectroscopy. Furthermore, \citet{Madhusudhan2020} showed that for K2-18~b, a H$_2$O layer beneath the atmosphere could have a surface as shallow as $\sim$1 bar and surface temperatures low enough for habitable liquid water. In this section, we investigate how thermal emission spectra of such planets can be used to learn about their atmospheres and to place limits on their surface conditions. In particular, we explore how observations using JWST/MIRI \citep{Rieke2015,Kalirai2018} can constrain both potentially habitable and warmer mini-Neptunes.

\subsection{Constraints on habitability}

In this section, we consider how atmospheric thermal emission observations can help to constrain the potential habitability of surface conditions in temperate mini-Neptunes similar to or smaller than K2-18~b (see section \ref{sec:intro}). In particular, thermal emission spectra can provide information about the temperature in the photosphere. The $P$-$T$ profiles in figures \ref{fig:PT_clouds} and \ref{fig:PT_main} show that the photospheres of these atmospheres can extend as deep as $\sim$0.1-1 bar, while high-altitude clouds can raise the photosphere to shallower pressures. Below the photosphere, different $P$-$T$ profiles can have a variety of gradients, from more isothermal profiles (e.g. due to ice clouds or haze) to relatively steep ones (e.g. due to high metallicity and/or low optical opacity), as demonstrated in figure \ref{fig:300}. Therefore, the photospheric temperature provides a lower limit to the temperature of a potential surface beneath the atmosphere. For example, if the brightness temperature of the atmosphere in an observed band is $>$400~K, the temperature at deeper pressures will be even higher, and habitable conditions are less likely. However, a photospheric temperature of $\sim$300~K could allow habitable conditions if the $P$-$T$ profile is isothermal below the photosphere, though a steeper temperature gradient could result in surface temperatures too high for habitability. For models with ice clouds, a high cloud opacity can result in the photosphere occurring at very low temperatures, e.g. $\lesssim$250 K, while making the $P$-$T$ profile isothermal below the cloud base and resulting in potentially habitable conditions at higher pressures. Observing such a low photospheric temperature, i.e. below the freezing point of H$_2$O, could therefore suggest the presence of such clouds and the possibility of a habitable surface.

For temperate mini-Neptunes with the potential for habitability, the emergent flux peaks in the mid-infrared ($\gtrsim$10 $\mu$m), and is therefore more easily observable at these wavelengths. Figures \ref{fig:spec_clouds} and \ref{fig:spec_main} show that the spectrum of a planet with irradiation temperature $\sim$350~K typically exceeds 10 ppm for wavelengths $\gtrsim$10 $\mu$m. Assuming optimistic uncertainties of $\sim$10 ppm, JWST observations in the mid-infrared (i.e. using MIRI) will be able to characterise these cool atmospheres and place limits on their habitability. Another advantage of this wavelength range is that the spectrum in the mid-infrared is predominantly sensitive to irradiation temperature, and is less sensitive to composition (see section \ref{sec:spectra}). This can be seen in figures \ref{fig:spec_clouds} and \ref{fig:spec_main}, where model spectra with different metallicities converge at longer wavelengths. The mid-infrared spectrum is therefore optimal for determining the temperature in the photosphere, regardless of its chemistry.

For a planet with mass and radius low enough for habitable surface pressures (e.g. K2-18~b), the photospheric temperature measured in the mid-infrared can therefore provide a quick metric for assessing whether a mini-Neptune could potentially host habitable conditions, and can rule out atmospheres which are likely too hot for habitability. Promising candidates can then be investigated further, e.g. using transmission spectroscopy to infer chemical compositions. Photometry provides an ideal way to measure this metric. In what follows, we explore how JWST/MIRI photometry can be used to measure photospheric temperature and distinguish potentially habitable planets from those which are too hot.

\subsection{JWST for mini-Neptune habitability}

JWST/MIRI will allow both photometric and spectroscopic observations beyond 10 $\mu$m, where potentially habitable mini-Neptunes can most easily be observed. Here, we assess how MIRI photometry can be used to constrain the photospheric temperatures of such planets. To do this, we simulate photometric data with the six MIRI photometric bands which probe wavelengths above 10 $\mu$m (F1130W-F2550W; \citealt{Rieke2015,Bouchet2015}) for a range of model atmospheres. The simulated photometric data are calculated by binning each model spectrum according to the instrument spectral response \citep{Glasse2015}. We assume best-case error bars of 10~ppm, which is a reasonable estimate for JWST photometry considering that the Hubble Space Telescope (HST) WFC3 spectrum of K2-18~b is known to a precision of $\sim$25~ppm at R$\sim$40 \citep{Benneke2019b,Tsiaras2019}. Figure \ref{fig:hab_obs} shows these data alongside the $P$-$T$ profiles for each model. For each photometric point, we also calculate the brightness temperature probed and indicate this on the corresponding $P$-$T$ profile; this gives an indication of the pressure being probed by each data point. For a spectral bin in the wavelength range $\lambda_{\mathrm{min}}$-$\lambda_{\mathrm{max}}$ and a normalised instrument sensitivity function $\zeta$, we calculate brightness temperature, $T_b$, such that 
\begin{equation*}
    \int_{\lambda_{\mathrm{min}}}^{\lambda_{\mathrm{max}}} \pi \zeta  B_{\lambda}(T_b)  d\lambda =  \frac{F_p}{F_s} \frac{R_s^2}{R_p^2}   \,\, \times \,\, \int_{\lambda_{\mathrm{min}}}^{\lambda_{\mathrm{max}}}  \pi \zeta I_s  d\lambda ,
\end{equation*}
where $\frac{F_p}{F_s}$ is the observed planet-star flux ratio in the photometric band, $\pi I_s$ is the stellar surface flux (we use Kurucz model spectra: \citet{Kurucz1979,Castelli2003}), $R_p$ is the planetary radius, $R_s$ is the stellar radius and $B_{\lambda}$ is the Planck function. 

Figure \ref{fig:hab_obs} shows the brightness temperatures corresponding to the six photometric bands and the pressures which they probe for each $P$-$T$ profile. In particular, for the model with ice clouds, the pressures probed are limited by the high optical depth of the cloud and are shallower than 0.01 bar. Conversely, for the warmest models with no ice clouds, the photometric points probe deeper than 0.1 bar. In all three cases, the brightness temperatures of the photometric points are cooler than the temperatures at higher pressures, where a surface could be expected to exist (e.g. $\gtrsim$1 bar, \citealt{Madhusudhan2020}). As such, the photometric data can provide a lower limit on the temperature of such a surface.

Figure \ref{fig:hab_obs} also shows that profiles whose photospheric temperatures differ by $\sim$100 K can easily be distinguished using MIRI photometry beyond 10 $\mu$m. Brightness temperatures $\gtrsim$300~K (e.g. for the warmer models in figure \ref{fig:hab_obs}) correspond to several tens - $\sim$100 ppm, which can be measured precisely if the error bars are $\sim$10 ppm. Brightness temperatures of $\sim$200-250 K (e.g. for the coolest model in figure \ref{fig:hab_obs}) can also be measured precisely using the longer-wavelength MIRI filters ($\gtrsim$17.5 $\mu$m), and can easily be distinguished from the warmer models. Indeed, the brightness temperature of this cooler model ($\sim$200~K) can be constrained to within $<$50 K using the F2550W filter (25.5~$\mu$m) and assuming 10 ppm uncertainties. As such, MIRI photometry provides a way to place lower limits on surface temperature and to rule out atmospheres which are likely too hot to host habitable surfaces. Furthermore, a non-detection with the longest-wavelength MIRI photometric band (F2550W) could suggest a very cool photospheric temperature and motivate further observations as habitable temperatures could be present deeper in the atmosphere. 

While we have used K2-18~b as a case study, these results also apply to similar planets around other M-dwarfs. Given the photospheric temperature of a planet, the observed flux ratio depends on the effective temperature of the star and is higher for cooler stars. Since K2-18 is an M2.5 V star, this case study provides a conservative estimate of the observability of mini-Neptunes orbiting M-dwarfs; later types can result in stronger signals.

\subsection{JWST for warmer mini-Neptunes}
\label{sec:obs_miri_steam}

Mid-infrared atmospheric observations can also be used to constrain the chemical and thermal properties of warmer mini-Neptunes. While these planets may not host habitable surfaces, their higher temperatures allow more detailed observations and constraints on their properties. In turn, such constraints can provide boundary conditions for internal structure models or give clues about their formation. Furthermore, since cooler mini-Neptunes are less suited to chemical characterisation with thermal emission, their warmer counterparts may help to constrain the dayside compositions of mini-Neptunes as a population. In this section, we investigate the observability of warm mini-Neptunes with JWST/MIRI using GJ~3470~b as a case study. 

To model GJ~3470~b, we use the bulk properties given in section \ref{sec:params} and spectroscopic atmospheric constraints from \citet{Benneke2019a}. \citet{Benneke2019a} find a significant detection of H$_2$O and constrain its abundance to be that expected from a solar metallicity atmosphere with an uncertainty of roughly $\pm 1$ dex. They also infer the presence of CO and/or CO$_2$ from the emission spectrum of the planet observed with Spitzer IRAC 1 and IRAC 2. However, since both of these species result in absorption features in the IRAC 2 4.5 $\mu$m bandpass, their abundances are degenerate with each other and form a knee-like degeneracy. We therefore do not vary the abundances of CO and CO$_2$ in our models and use the median mixing ratios inferred by \citet{Benneke2019a}, i.e. $10^{-3.04}$ and $10^{-2.71}$, respectively, as the enhancement of one would merely lead to the depletion of the other. \citet{Benneke2019a} further infer a depletion of CH$_4$ and NH$_3$ in GJ~3470~b, so we omit these species from our models. Since the abundances of CO and CO$_2$ relative to H$_2$O are significantly higher than expected from equilibrium chemistry, we choose to use constant-with-depth abundances for these three species. We also include KCl and ZnS clouds as described in section \ref{sec:model}, with cloud bases at 0.3~bar. The $P$-$T$ profiles and spectra for these models are shown in figure \ref{fig:steamy_obs}. As in figure \ref{fig:hab_obs}, we also show simulated MIRI photometry assuming uncertainties of 10~ppm, which allows high signal-to-noise measurements of the emergent flux and photospheric temperature.

Figure \ref{fig:spec_main} shows that a model spectrum with $T_\mathrm{irr}$=800~K exceeds an optimistic uncertainty of 10 ppm at wavelengths greater than $\sim$3 $\mu$m. In this wavelength range, the MIRI Low Resolution Spectroscopy (LRS) and Medium Resolution Spectroscopy (MRS) can be used to observe the thermal emission from these planets. In figure \ref{fig:steamy_obs}, we show simulated MIRI LRS data for the model spectra of GJ~3470~b. We simulate this data using PandExo \citep{Pandexo} for the MIRI LRS slitless mode. For the error bars we assume a single eclipse, and the data are binned by a factor of 10. Absorption features due to H$_2$O, CH$_4$, CO$_2$ and HCN are present in this spectral range and, using retrieval analyses, abundances of these species could be inferred from such spectra. For example, a strong absorption feature is visible at $\sim$9-10 $\mu$m. In addition to chemical information, the emission spectrum can also be used to constrain the $P$-$T$ profile of the planet. In the examples shown in figure \ref{fig:steamy_obs}, the spectra probe pressures within the range $\sim$0.01-0.1~bar, meaning that the temperature profile in this range could be inferred using retrieval techniques.

\section{Discussion \& Conclusions}
\label{sec:discussion}
Mini-Neptunes in the habitable zones of M-dwarfs can provide an excellent opportunity to study temperate exoplanets and their potential habitability. In particular, \citet{Madhusudhan2020} showed that the mass, radius and atmospheric properties of the habitable-zone mini-Neptune K2-18~b allow it to have a H$_2$O surface beneath the atmosphere with a surface pressure as low as $\sim$1~bar, potentially with habitable conditions. In this study, we explore the effects of a range of atmospheric parameters on the thermal profiles and spectra of such planets. We investigate three primary aspects of mini-Neptune atmospheres: (i) the diversity of atmospheric temperature structures, with implications for energy transport, chemistry and their emergent spectra, (ii) thermodynamic conditions deep in the atmosphere, which impact the potential for habitability as well as boundary conditions for internal structure models, and (iii) their observability with thermal emission spectra, including constraints on potential habitability.

We begin by exploring the diversity of temperature structures and emission spectra of mini-Neptunes as a function of several atmospheric parameters. The parameters we consider are irradiation temperature, internal temperature, metallicity and cloud/haze properties. We find that for typical internal temperatures of mini-Neptunes, the radiative-convective boundary occurs well below the photosphere. As a result, vertical mixing in the photosphere is more likely to be caused by eddy mixing rather than convection. We also find that strong optical opacity due to clouds/hazes can result in an isothermal temperature structure beneath the photosphere, which can maintain cool temperatures to high pressures, with implications for habitability on deep surfaces. For all of the models we consider, which have irradiation temperatures $\geq$350~K, we further find that the emergent spectra are above the 10~ppm level at wavelengths $>$10~$\mu$m. This means that the emission spectra of such mini-Neptunes could be observable with JWST/MIRI, which we discuss below.

We also apply our atmospheric model to the habitable-zone mini-Neptune K2-18~b to asses its atmospheric conditions and potential habitability. We consider a range of physically-motivated thermal profiles by varying metallicity, internal temperature and haze abundance. We compare our $P$-$T$ profiles to the phase diagram of water in order to assess the phase that a H$_2$O surface would have given a particular surface pressure. We find that many of our models intersect the liquid or supercritical phases at pressures $\gtrsim$1~bar, suggesting that liquid or supercritical surface water could be possible on this planet for a range of atmospheric conditions. This has implications for internal structure models of mini-Neptunes as the equation of state of H$_2$O is strongly temperature-dependent for phases other than ice \citep{thomas2016,Madhusudhan2020}.

We also explore atmospheric models for K2-18~b which allow liquid surface water at habitable temperatures and pressures, i.e. $T\lesssim$395~K and $P\lesssim$1250~bar, corresponding to the most extreme temperature and pressure conditions habitable for extremophiles on Earth \citep{Merino2019}. \citet{Madhusudhan2020} model the atmosphere and interior of K2-18~b and find that, for two atmospheric $P$-$T$ profiles, solutions with habitable liquid water at the surface are possible. Here, we explore a variety of atmospheric models for K2-18~b spanning a range of metallicities, internal temperatures and haze opacities and find several solutions which allow habitable liquid water at the surface. We also consider a hypothetical planet resembling K2-18~b but with $T_\mathrm{irr}$=420~K rather than 332~K (i.e., receiving $\sim$2.5$\times$ more incident flux). We find that, despite its hotter irradiation temperature, several model solutions for this planet also allow liquid surface water at habitable temperatures and pressures. In these scenarios, stronger optical scattering is needed to sufficiently cool a planetary surface. These results suggest that mini-Neptunes throughout the traditionally-defined habitable zone could potentially host habitable conditions, depending on their atmospheric properties and surface pressures.

In our current models, optical scattering due to clouds/hazes is necessary to cool the lower atmosphere to habitable temperatures. In principle, some of the dayside energy may also be redistributed to the nightside, thereby cooling the dayside atmosphere. Here, we have conservatively assumed that any redistribution of the stellar irradiation from the dayside to the nightside occurs in the interior, below the atmosphere. In principle, where the interior-atmosphere boundary is deep enough, the day-night energy redistribution can occur in the atmosphere, as is known to occur in irradiated gas giants \citep{Showman2009}. In the context of 1-D models, such an effect can be modeled by an artificial energy sink in the dayside atmosphere \citep[e.g.][]{Burrows2008a}. If energy is removed from the day-side at pressures $\lesssim1000$~bar, this could contribute to cooling the atmosphere at habitable pressures, thereby increasing the chances of habitable conditions in our models. As such, our estimates for habitable conditions in the present work may be conservative. Including such an effect may also mean that less optical opacity is required for habitable conditions to be possible. This can be explored in future work, alongside more complex prescriptions for clouds and hazes, in order to understand the potential habitability of mini-Neptunes in greater detail.

We further note that our current models do not include the effects of water condensation on convection and the temperature profile. Latent heat released by the condensation of water is known to make convection easier in some circumstances \citep[e.g.][]{Pierrehumbert2010}, resulting in a more isothermal `moist' adiabat. Conversely, in atmospheres where the background gas is lighter than the condensing species (e.g. H$_2$O condensation in a H$_2$-rich atmosphere), condensation can inhibit both moist and double-diffusive convection and result in a super-adiabatic temperature gradient \citep[e.g.][]{Guillot1995,Leconte2017,Friedson2017}. The ways in which these effects can shape the energy transport and $P$-$T$ profiles in H$_2$-rich atmospheres are not yet fully understood, though future missions to Uranus or Neptune may help to elucidate this \cite{Guillot2019}.

Finally, we consider the observability of mini-Neptune thermal emission with JWST, including observable signatures of potentially habitable conditions. We find that photometry with JWST/MIRI could be used to measure the photospheric temperatures of mini-Neptunes such as K2-18~b down to $\sim$200~K assuming 10~ppm error bars. Based on our models, we find that photospheric temperatures below the freezing point of H$_2$O suggest the presence of ice clouds and could be a sign of temperate conditions below the atmosphere. Photospheric temperatures of $\sim$300-400~K can also be a sign of habitable conditions below the atmosphere, though this depends on the temperature gradient at higher pressures. This gradient is driven by several factors including infrared opacity and the presence of hazes/clouds. However, photospheric temperatures $\gtrsim$400~K typically imply even hotter, and therefore non-habitable, conditions at higher pressures. 

Our models therefore show that MIRI photometry can provide a simple way to establish whether a mini-Neptune such as K2-18~b could potentially host habitable conditions, guiding follow-up observations. For example, transmission spectroscopy could provide more detailed chemical constraints which in turn could be used to provide further insight into the habitability of the planet \citep[e.g.][]{Seager2013b,Bains2014,Meadows2018book}. In addition to temperate, potentially-habitable mini-Neptunes, we also explore the observability of warmer mini-Neptunes with JWST and find that MIRI low resolution spectroscopy could be used to constrain their chemical compositions and thermal profiles.

Our results show that mini-Neptunes similar to or smaller than K2-18~b, whose masses and radii allow for habitable surface pressures, could potentially host habitable conditions beneath their H$_2$-rich envelopes for a wide range of atmospheric parameters. Since these planets are more easily observable than temperate terrestrial planets, they arguably represent optimal targets for the study of habitability in exoplanets. Furthermore, JWST/MIRI could provide initial constraints on the habitability of such mini-Neptunes.

\acknowledgments
{AAAP acknowledges support from the UK Science and Technology Facilities Council (STFC) towards her doctoral studies. We thank the anonymous reviewer for their helpful review of our work. This research has made use of the NASA Astrophysics Data System and the Python packages \textsc{numpy}, \textsc{scipy} and \textsc{matplotlib}.}

\bibliography{refs}{}
\bibliographystyle{aasjournal}



\end{document}